\documentclass[onecolumn,11pt]{article}
\usepackage[top=1in, bottom=1in, left=1in, right=1in]{geometry}
\usepackage{palatino} 
\usepackage{xcolor,fullpage}
\usepackage{newtxtext}
\usepackage{mathrsfs}
\usepackage{graphicx}
\usepackage{fancyhdr}
\usepackage{amsmath}
\usepackage{amssymb}
\usepackage{textcomp}
\usepackage[utf8]{inputenc}
\usepackage[square,sort,comma,numbers]{natbib}
\usepackage{subcaption}	
\usepackage{hyperref}
\usepackage{setspace}
\usepackage{cancel}
\usepackage{mathtools}
\usepackage[font=footnotesize]{caption}
\usepackage{rotating}
\usepackage{adjustbox}
\usepackage{multirow}
\usepackage{tabularx}
\usepackage{lipsum}
\usepackage{booktabs}
\usepackage{subfiles}
\usepackage{titling}
\usepackage{placeins}
\usepackage{authblk}
\usepackage{comment}
\usepackage{wrapfig}
\usepackage{chngcntr}
\usepackage[normalem]{ulem}
\usepackage{outlines}
\usepackage{mathabx}
\usepackage{makecell}   
\usepackage{enumitem}

\newcolumntype{L}[1]{>{\raggedright\let\newline\\\arraybackslash\hspace{0pt}}m{#1}}
\newcolumntype{C}[1]{>{\centering\let\newline\\\arraybackslash\hspace{0pt}}m{#1}}
\newcolumntype{R}[1]{>{\raggedleft\let\newline\\\arraybackslash\hspace{0pt}}m{#1}}

\newcounter{SIsection}
\renewcommand{\theSIsection}{SI \arabic{SIsection}}

\newcommand{\SIsection}[1]{%
    \refstepcounter{SIsection}%
    \section*{\theSIsection \hspace{1em} #1}%
}

\usepackage{lineno} 

\title{\vspace{-.5in}\textbf{\huge{Weak Dominant Balance for Robust Identification of Dynamically Consistent Fluid Flow Structure}}\vspace{-.25in}}%

\author[1,2]{Samuel Ahnert$^*$}
\author[1,2]{Esther Lagemann}
\author[3]{H. Jane Bae}
\author[4]{Kunihiko Taira}
\author[5]{Ricardo Vinuesa}
\author[1,2]{Christian Lagemann}
\author[1,2]{Steven L. Brunton$^*$}

\affil[1]{\small Department of Mechanical Engineering, University of Washington, Seattle, WA, USA}
\affil[2]{\small AI Institute in Dynamic Systems, University of Washington, Seattle, WA, USA}
\affil[3]{\small Lynn Booth and Kent Kresa Department of Aerospace, California Institute of Technology, Pasadena, CA, USA}
\affil[4]{\small Department of Mechanical and Aerospace Engineering, University of California, Los Angeles, CA, USA}
\affil[5]{\small Department of Aerospace Engineering, University of Michigan, Ann Arbor, MI, USA}
\affil[$*$]{{\footnotesize corresponding authors (sahnert@uw.edu, sbrunton@uw.edu)}}

\begin{document}
\date{}
\vspace{-0.75in}
\maketitle
\vspace{-0.85in}
\begin{abstract}
\normalsize
Extracting interpretable, localized physical mechanisms from complex spatiotemporal data is a foundational challenge across physics, biology, and engineering, but has remained out of reach on real measurements. 
The central obstacle is obtaining high-quality gradients of data via numerical differentiation, which amplifies noise, diverges for high-order equations, and falters on irregular geometries, limiting the scope of existing approaches to clean simulations of low-order systems.
Here, we present \emph{weak dominant balance}, a derivative-free framework that projects governing equations into a weak (integral) formulation, offloading differentiation onto smooth analytical test functions and leaving the data untouched. 
The method sustains accurate regime identification under severe noise where existing approaches categorically fail, 
delivers the first data-driven decomposition of a third-order partial differential equation applied to turbulent duct flow, 
and produces matching decompositions across direct numerical simulation and particle-image velocimetry measurements of a wavy channel flow, uncovering a previously uncharacterized dynamical regime.
Weak dominant balance brings mechanism-level analysis out of simulation and onto measured data, and opens complex physical systems to direct, equation-grounded interpretation.
\end{abstract}
\vspace{-.2in}

\section{Introduction}
\label{Sec: Intro}

Reducing complex physical systems to key localized governing mechanisms has driven progress in science and engineering for over a century. In fluid dynamics, this paradigm is exemplified by Prandtl's boundary-layer asymptotics~\cite{prandtl1928motion} and Kolmogorov's similarity arguments~\cite{K41_1941}, which simplify the full Navier--Stokes equations to distinct, region-specific balance models. While modern modal-decomposition techniques~\cite{JLumley_FirstFluidsPODPaper, schmid2002_GlobalStability, ROWLEY_DMD, Haller_LCS_2002, McKEON_SHARMA_2010} and machine-learning extensions~\cite{ArranzLozanoDuran_IND_For_Turbulent_Flows, SHAP_Vinuesa, vinuesa2026explainableailearninglearners} excel at extracting coherent patterns directly from observed fields, they primarily operate agnostic to the governing equations themselves, frequently decoupling the identified structures from the underlying mathematical physics. 
Restoring this connection requires an analysis whose coordinates are explicitly grounded in the individual terms of the governing partial differential equation (PDE)---an ``equation-space'' in which unsupervised clustering can systematically partition a physical domain into distinct, dynamically consistent regimes~\cite{CallahamDB,OceanicDynamicsDB_2019}.

\begin{figure}[t!]
    \begin{center}
        \includegraphics[width=\textwidth]{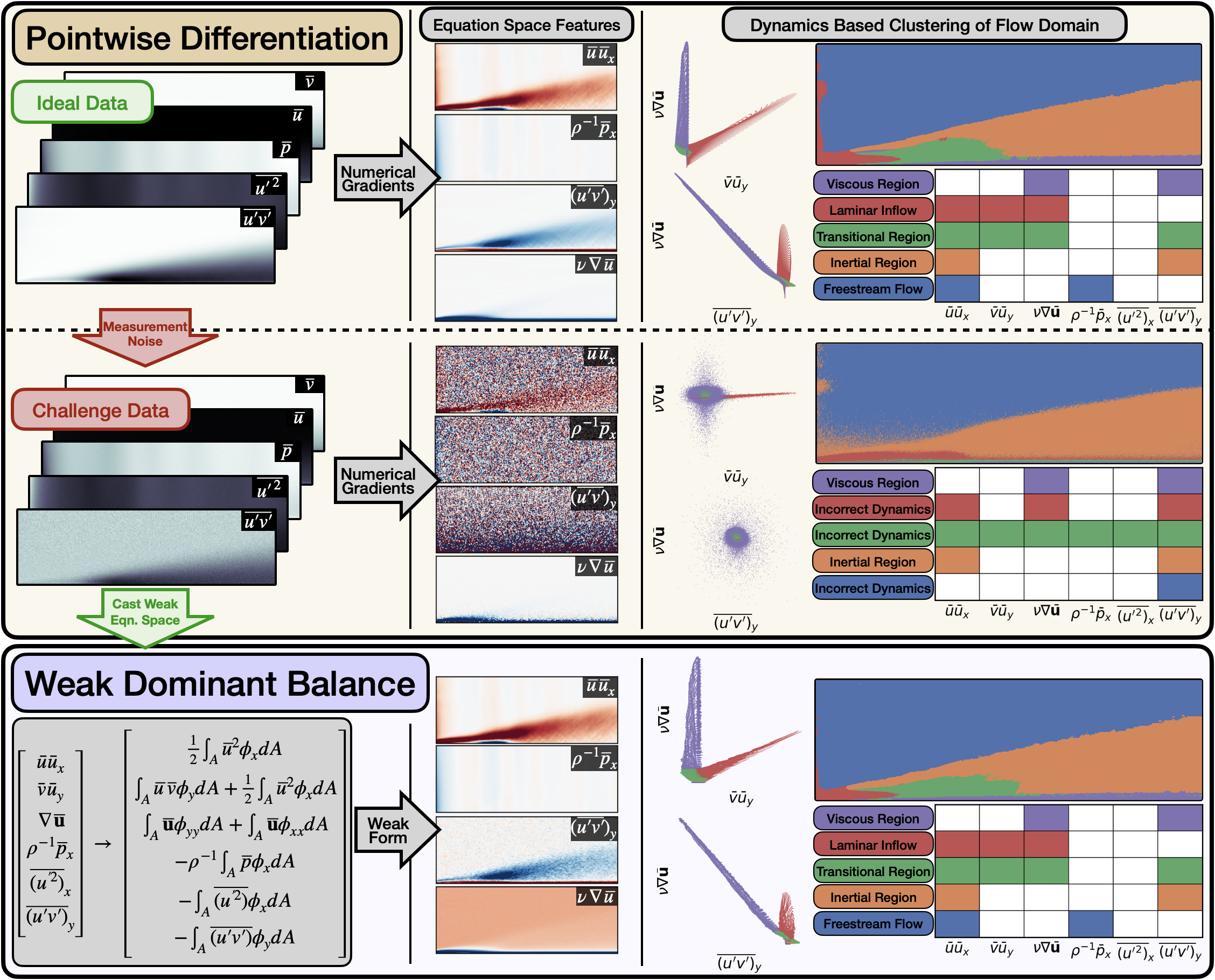}
    \end{center}
    \caption{\textbf{Weak dominant balance enables equation-space regime identification on noisy data.}
        The three rows analyze the same transitional flat-plate turbulent boundary layer. Only the data quality and the mathematical representation of the equation-space coordinates differ between rows. \textbf{Top:} Reference decomposition computed from noise-free fields using pointwise numerical differentiation as in~\cite{CallahamDB}. The equation-space embedding (center) is well-structured, and unsupervised clustering of the embedding partitions the spatial domain (right) into the five canonical regimes of a transitional boundary layer---freestream, laminar inflow, transitional, inertial, and viscous. The sparse-PCA dynamics matrix on the far right shows which terms of the streamwise Reynolds-averaged Navier--Stokes equation are active in each regime. \textbf{Middle:} The same pipeline applied to fields contaminated with additive Gaussian noise ($10\%$ noise as defined in Sec.~\ref{sec: TBL_NoiseStudy}). Pointwise differentiation amplifies high-frequency noise across every gradient term, collapsing the equation-space manifold and rendering the resulting domain partition physically meaningless: three of the five regimes (red labels) are misidentified. \textbf{Bottom:} Under identical noise, the weak formulation computes each equation-space coordinate as the inner product of a measured field against an analytically differentiated test function $\phi$ (boxed equations, left). Derivatives are evaluated on smooth analytical objects rather than on discrete data, preserving the embedding structure and recovering the reference regimes with high fidelity. The transition from pointwise to weak construction acts only on how the equation-space coordinates are computed; the downstream clustering and term-identification pipeline is identical across all three rows.}
\label{Fig:Overview}
\end{figure}

Realizing this idea on real-world measurements, however, has remained out of reach given the severe computational and mathematical bottleneck of evaluating equation-space coordinates. In particular, obtaining accurate spatial gradients of discrete, sampled fields is a major obstacle that is highly sensitive to measurement noise, which numerical differentiation amplifies progressively at higher derivative orders. Furthermore, finite-difference stencils are challenging to implement on the unstructured, non-rectilinear, or curvilinear meshes typical of real-world geometries, and they incur severe computational cost if evaluated over large domains. Consequently, equation-space clustering has been largely restricted to clean, idealized direct numerical simulations (DNS) of low-order, two-dimensional systems on structured grids, excluding the very systems where automated regime identification would be most valuable: experimental measurements, complex boundaries, and higher-order dynamics.

Here, we present \emph{weak dominant balance}, a derivative-free framework that constructs equation-space coordinates as inner products of each governing-equation term against a dense bank of smooth, compactly supported test functions. Performing integration by parts mathematically transfers spatial derivative operators from the discrete, noisy data onto the test functions, where they can be evaluated analytically. For two-dimensional incompressible flows, we further exploit incompressibility to recast the advective terms into a conservation form, completely eliminating spatial derivatives of the velocity field from the equation-space construction. The result is accurate, equation-grounded regime discovery coming directly from noisy, high-order, and geometrically complex observations.

Figure~\ref{Fig:Overview} illustrates the dramatic shift in results when opting to apply weak dominant balance on a transitional flat-plate turbulent boundary layer compared to a pointwise (finite-difference) baseline. In the absence of noise, both formulations yield structured equation-space embeddings and partition the domain into expected physical regimes (Fig.~\ref{Fig:Overview}, top). However, when modest measurement noise is introduced, pointwise numerical differentiation amplifies the noise across all gradient terms, collapsing the equation-space manifold and rendering the resulting classification physically meaningless (Fig.~\ref{Fig:Overview}, middle). In contrast, our weak formulation transfers derivatives to analytical test functions, preserving the structure of the equation-space embedding and successfully recovering the reference physical regimes even under severe noise levels (Fig.~\ref{Fig:Overview}, bottom). Beyond noise robustness, we later demonstrate that this mathematical restructuring of the equation space generalizes to high-order PDEs and complex experimental geometries. Overall, this work makes the following contributions:
\begin{enumerate}[leftmargin=*,itemsep=1pt,topsep=2pt]
  \item \textbf{Derivative-free equation-space analysis:} A weak formulation that constructs equation-space coordinates as integral projections against compactly supported test functions, transferring differentiation onto analytical objects and admitting analysis on noisy, high-order, and geometrically irregular data (Sec.~\ref{sec: WeakDBAlgoOverview}).
  \item \textbf{Conservative Advection Formulation:} We derive a mathematically consistent, conservation-form representation of the 2D incompressible Reynolds-Averaged Navier--Stokes (RANS) equations, eliminating all velocity gradients from the advective terms in the equation-space construction (Sec.~\ref{sec: 2DIncompressibleRANS}).
  \item \textbf{Noise-Robustness Benchmarks:} We demonstrate through systematic noise studies on a transitional boundary layer that the weak framework sustains a median classification error below $2.71\%$ up to $100\%$ noise, whereas pointwise methods degrade immediately at $1\%$ noise (Sec.~\ref{sec: TBL_NoiseStudy}).
  \item \textbf{High-Order and Experimental Demands:} We present the first data-driven dominant-balance analysis of a third-order PDE---the vorticity transport equation---isolating the specific Reynolds-stress mechanisms that generate corner vortices in a turbulent duct (Sec.~\ref{sec:duct_vorticity}). Furthermore, using matched DNS and experimental particle-image velocimetry (PIV) data of a turbulent wavy channel, we show that the weak framework recovers nearly identical dynamical decompositions under both settings and identifies a previously uncharacterized physical regime in the wave trough (Sec.~\ref{Sec:WavyWallResults}). These analyses are all shown to compute at comparable or significantly faster wall clock speeds than a pointwise baseline by the efficient implementation of the weak approach in Fourier space (Sec.~\ref{sec:FFTSpeedDemon}).
\end{enumerate}

Although our demonstrations are drawn from fluid mechanics, the framework is physics-agnostic. It applies to any spatiotemporal system where the candidate governing PDE is known and the physical state can be observed. Potential targets across physical and biological sciences include the multiscale, multiphase processes in the cardiovascular system~\cite{quarteroni2017integrated, Quarteroni_Manzoni_Vergara_2017}, 
sparse multi-mechanism balances in geophysical modelling~\cite{Vallis2016_GFDPerspective, SALCEDOSANZ2020256, bodnar2025foundation}, advances in theoretical understanding of oceanic and ecological dynamics~\cite{Sonnewald_2021, EcologicallySimilarProvinces_2020, OceanicDynamicsDB_2019, Global_Heating_2021},
regimes and instabilities of magnetohydrodynamic flows in fusion plasmas~\cite{park2025kinetic, Kaptanoglu:2020vu}, 
and the complex reactive--convective--diffusive balances in a flame front~\cite{zeng2020complex}.

\section{Background}
This section provides an overview on the mathematical background that is required to construct the weak dominant balance method. Equation-space analysis is used as a general framework for partitioning data from physical systems into dynamically consistent regimes (Sec.~\ref{Sec:DominantBalance}), and we describe the weak formulation of differential operators (Sec.~\ref{Sec:WeakSINDy}) that is later exploited in Sec.~\ref{Sec:Methods} to lift this framework from idealized simulations onto noisy, high-order, and geometrically complex measurements.

\subsection{Equation-space analysis of governing PDEs}\label{Sec:DominantBalance}

The methodology we develop in this work belongs to a broader class of approaches that analyze the diversity of behaviors in chaotic, multi-scale systems through mechanisms in the equations themselves~\cite{BuckinghamPi, Sterrett_PhysicallySimilarSystems, Advanced_Mathematical_Methods_for_Scientists_and_Engineers}. %
Specifically, the locally dominant physical processes may be discovered in \emph{equation space}---a representation in which each coordinate corresponds to an individual term of the governing partial differential equation. For a PDE that can be written as a sum of linearly separable terms,
\begin{equation}
\mathcal{N}(\mathbf{u}) \;=\; \sum_{i=1}^{k} f_i(\mathbf{u}) \;=\; 0,
\label{eq:pde_general}
\end{equation}
the equation-space embedding maps each spatial point to the vector
\begin{equation}
\mathbf{V} \;=\; \bigl[\, f_1(\mathbf{u}),\, f_2(\mathbf{u}),\, \ldots,\, f_k(\mathbf{u}) \,\bigr]^T,
\label{eq:eqn_space}
\end{equation}
so that data points governed by common physical mechanisms cluster into manifolds in $\mathbb{R}^k$ that can be partitioned by unsupervised learning. This perspective formalizes the classical practice of identifying dominant balances---Prandtl's boundary-layer scaling~\cite{prandtl1928motion}, Kolmogorov's inertial-range arguments~\cite{K41_1941}, von K\'arm\'an's Law of the Wall~\cite{Karman1930_LawOfTheWall}, Stommel's western-boundary-current asymptotics~\cite{Strommel_OceanCurrentIntensification}---and removes the requirement that the practitioner specify the active terms in advance. A concrete realization of this idea, using finite-differencing to set up equation-space, Gaussian mixture models \cite{Bishop_GMM_2006} for clustering, and sparse principal component analysis \cite{Zou_sPCA_2006} for term identification, was introduced by Callaham et~al.~\cite{CallahamDB} and has since been applied to canonical second-order PDEs on structured grids (see \ref{SI:DominantBalance} for more details).

\subsection{Weak Calculation of Partial Derivatives}
\label{Sec:WeakSINDy}
Data-driven discovery of dynamical systems often requires the calculation of derivatives from noisy measurement data~\cite{Brunton2016pnas,Rudy2017sciadv}. This becomes particularly challenging in the context of learning PDEs~\cite{Rudy2017sciadv}, where fields that contain high-order partial derivatives become contaminated by noise.  
Several advances have been made to improve these computations, notably, by incorporating ensembling techniques~\cite{EnsembleSINDy}, or by recasting the problem in a weak integral formulation~\cite{IntegralSINDy,WeakSINDY_GalerkinBasedODERegression,WSINDyForPDEs,Grigoriev_HisFirstWSINDyPaper,bioinformatics_WSINDy,WANG201944_WSINDy}.  
We focus on the weak formulation, which has enabled learning from orders of magnitude more measurement noise and is directly relevant for the computation of partial derivatives required for dominant balance analysis. Contrary to the goals in dynamics regression, the governing PDE in dominant balance is known a priori. Accordingly, we can convert an example governing equation $\dot{x}(t) = f'(x(t))$ into its weak form, where $x(t)$ represents observations of the state over time and $f'(x(t))$ is a function of the data that includes a spatial derivative for demonstration purposes. 
Given a test function $\phi$ compactly supported over an interval $[a,b]$ contained within the domain of the data, the weak form of this PDE may be written:
\begin{equation}
    \label{eqn: weakFormOFDynamics_unsimplified}
    \int_a^b \phi \dot{x}(t) dx = \int_a^b \phi  f'(x(t)) dx.
\end{equation}
Applying integration by parts to the right hand side of Eq. (\ref{eqn: weakFormOFDynamics_unsimplified}) allows us to offload the spatial derivative from a data field to the test function as follows:
\begin{equation}
    \label{eqn: weakFormOFDynamics_expanded}
    \int_a^b \phi \dot{x}(t) dx = [\phi f(x(t)]_a^b-\int_a^b \phi'  f(x(t)) dx.
\end{equation}
Application of the compact support of $\phi$ on $(a,b)$ reduces Eq. (\ref{eqn: weakFormOFDynamics_expanded}) to:
\begin{equation}
    \label{eqn: weakFormOFDynamics_simplified}
    \int_a^b \phi \dot{x}(t) dx = -\int_a^b \phi'  f(x(t)) dx,
\end{equation}

effectively eliminating the need for spatial derivatives of the data to reconstruct our right-hand side term, and this process applies for an arbitrary number of linearly separated terms containing spatial gradients.

By casting a weak formulation, we gain two primary advantages for equation space analysis. First, derivatives are evaluated on smooth analytical objects rather than on discrete data, eliminating the dominant source of error when differentiating noisy fields. Second, because the inner products in Eq.~\eqref{eqn: weakFormOFDynamics_simplified} reduce to convolutions in Fourier space~\cite{WSINDyForPDEs}, the equation-space coordinates can be computed in a single FFT pass which in practice is orders of magnitude faster than assembling and applying finite-difference operators of comparable spatial extent (see Section~\ref{sec:FFTSpeedDemon} for wall-clock comparisons). 

\section{Methods} \label{Sec:Methods}


The premise of equation-space analysis is that the coordinates $f_i(\mathbf{u})$ of the embedding in Eq.~\eqref{eq:eqn_space} accurately represent the local magnitude of each physical mechanism in the governing PDE. Whether those coordinates can be constructed at all on a given dataset, and whether they remain meaningful under realistic measurement conditions, is therefore the central methodological question. In this section we show that recasting the equation-space construction in a weak (integral) form resolves three obstacles simultaneously: sensitivity to measurement noise, breakdown on high-order derivative terms, and incompatibility with non-rectilinear geometries. In Sec.~\ref{sec: WeakDBAlgoOverview} we derive the general weak equation-space coordinates, and in Sec.~\ref{sec: 2DIncompressibleRANS} we show that for two-dimensional incompressible flow the construction can be made
\emph{exactly} derivative-free.

\subsection{Weak Dominant Balance}
\label{sec: WeakDBAlgoOverview}

In order to cast a weak form of the governing equation, we construct a grid of test functions $\Phi = \{\phi_1, \dots, \phi_n\}$ that densely populates the data domain, $\Omega$, each of which is a translation of a single reference function $\phi_r$. We take $\phi_r \in C_c^p(\Omega)$, smooth and compactly supported on $\Omega$, using the family of piecewise polynomial functions defined in~\cite{WSINDyForPDEs},
\begin{equation}
  \phi_r(\mathbf{x}) \;=\;
  \begin{cases}
  \prod_{i=1}^{d}\, \bigl(1 - (x_i/L_i)^{2}\bigr)^{\,p} & -L_i<x_i<L_i \\
  0, & \text{otherwise},
  \end{cases}
\end{equation}
with support half-width $L_i$ in each spatial direction and polynomial order $p$. The order $p$ must satisfy $p \ge m$ in order to avoid trivial solutions, where $m$ is the maximum derivative order applied to $\phi_r$ in the weak equation-space coordinate transformation (Eq.~\eqref{eq: WeakRANS_2D_CompactForm}). For the streamwise RANS case this requires $p \ge 2$, and for the vorticity-transport case $p \ge 2$ after the reduction in Eq.~\eqref{eq: WeakViscousVorticity}, although in practice, $p$ must be substantially higher than the minimum required for integration by parts as noted in~\cite{WSINDyForPDEs,Bortz2023_WENDY_BumpFunctions}.

\begin{figure}[t!]
    \begin{center}
        \includegraphics[width=\textwidth]{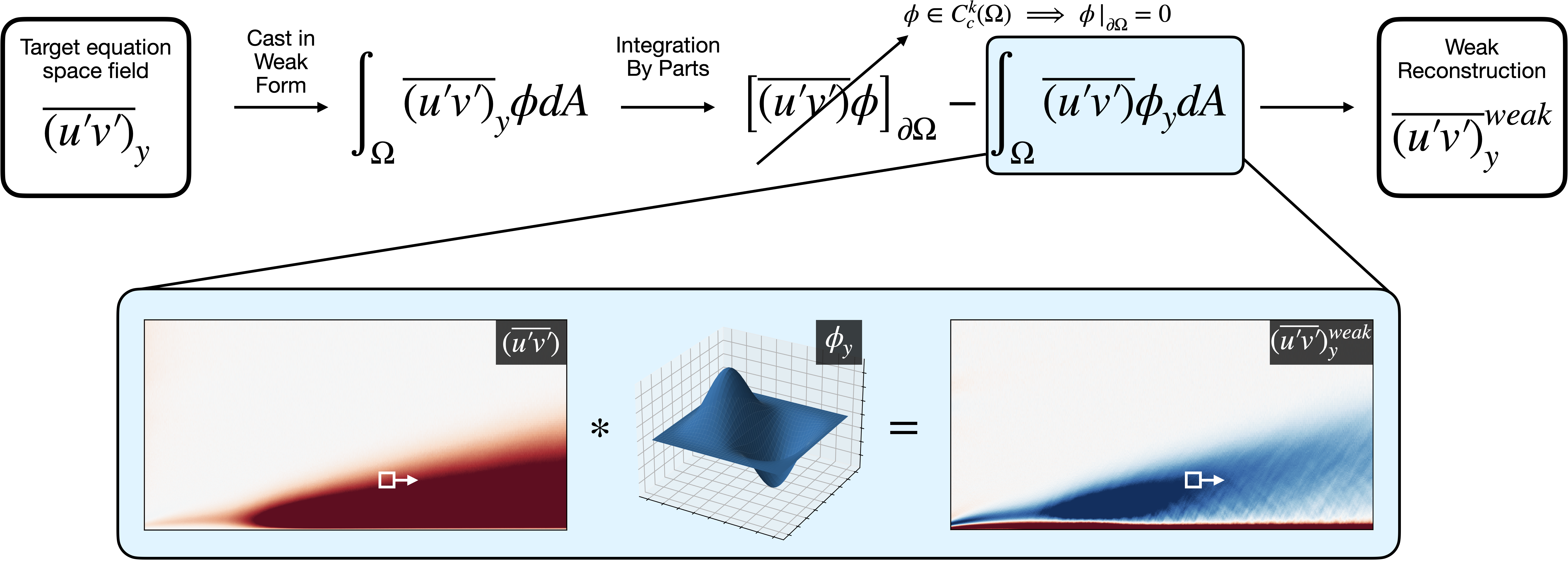}
    \end{center}
    \caption{\textbf{Computing a weak equation-space coordinate by convolution with an analytically differentiated test function.} \textbf{Top row (Derivation):} The target equation-space term, the Reynolds-shear-stress gradient $\overline{u'v'}_y$, is recast in weak form by multiplying with a compactly supported test function $\phi \in C^k_c(\Omega)$ and integrating. A single integration by parts transfers the spatial derivative from the measured field $\overline{u'v'}$ onto the test function, producing $-\!\int_\Omega \overline{u'v'}\,\phi_y\, dA$; the boundary term vanishes by the compact support of $\phi$. The result is a weak reconstruction $(\overline{u'v'}_y)^{\text{weak}}$ that contains no numerical derivative of the data. \textbf{Bottom row (Evaluation):} Translating $\phi_y$ across the domain and evaluating the inner product at each center point is mathematically equivalent to a single convolution of the data field $\overline{u'v'}$ with the analytically differentiated reference test function $\phi_y$---which can be computed in one FFT pass, regardless of the derivative order embedded in $\phi$. This same construction applies term by term to every coordinate of the equation-space embedding, replacing the entire finite-difference stencil bank of pointwise dominant balance with a bank of analytical test functions.}
\label{fig: Fig02_Methods}
\end{figure}

Let us consider the streamwise incompressible RANS equation (defining the $x$-direction to be streamwise going forward) for a dataset consisting of mean velocities $\overline{U}, \overline{V}, \overline{W}$ in the $x,y,z$ directions, mean pressure field $\overline{p}$, and Reynolds stresses $\overline{u'^2}, \overline{u'v'}, \overline{u'w'}$, 
\begin{equation}
\overline{U} \ \overline{U}_x + \overline{V} \ \overline{U}_y + \overline{W} \ \overline{U}_z = -\frac{\overline{p}_x}{\rho} + \nu(\overline{U}_{xx} + \overline{U}_{yy} + \overline{U}_{zz}) - (\overline{u'^2}_x + \overline{u'v'}_y + \overline{u'w'}_z).
\label{eqn: RANS} 
\end{equation}
To cast a weak form of the governing equation, we multiply by our grid of test functions $\Phi$ and integrate, which is typically done to bypass the stringent criteria of smoothness on the solutions for a system of interest. Take the shear Reynolds stress term in Eq. (\ref{eqn: RANS}) as an example, the weak reconstruction of this term at a single point at the center of a single $\phi \in \Phi$ for simplicity can be written:
\begin{equation}
\int_\Omega (\overline{u'v'}_y) \space \phi \space d\Omega.
\label{eqn: weak Reconstruction of Normal RS} 
\end{equation}
Performing integration by parts on Eq. (\ref{eqn: weak Reconstruction of Normal RS}) now allows us to offload the $y$-gradient from data onto the test function as follows,
\begin{equation}
\int_\Omega (\overline{u'v'}_y) \space \phi \space d\Omega = \left[(\overline{u'v'}) \space \phi\right]_{\partial\Omega} -\int_\Omega (\overline{u'v'}) \space \phi_y \space d\Omega,
\label{eqn: IBP Normal RS} 
\end{equation}
where the resulting $[(\overline{u'v'} )\phi]_{\partial\Omega}$ is definitionally zero by the compact support of $\phi$ on $\Omega$, giving our final reconstruction with no derivative on the $\overline{u'v'}$ data field. In Fig.~\ref{fig: Fig02_Methods}, we outline the full derivation of a term $(\overline{u'v'})_y^{weak}$, demonstrating the convolution approach to reconstruction using a once differentiated reference test function, $\phi_y$, and the data field $\overline{u'v'}$, in order to offer an intuitive visualization of the typical workflow for weak form scientific machine learning methods.

Repeating this process for each term in the streamwise RANS equation (\ref{eqn: RANS}), we obtain the following full weak form:
\begin{equation}
\label{eqn: WeakRANSFullEqn} 
    \begin{gathered}
        \int_\Omega (\overline{U} \ \overline{U}_x) \space \phi \space d\Omega + 
        \int_\Omega (\overline{V} \ \overline{U}_y) \space \phi \space d\Omega + 
        \int_\Omega (\overline{W} \ \overline{U}_z) \space \phi \space d\Omega = 
        \int_\Omega \frac{\overline{p}}{\rho} \space \phi_x \space d\Omega + 
        \nu (
        \int_\Omega \overline{U} \space \phi_{xx} \space d\Omega
        \\
        +
        \int_\Omega \overline{U} \space \phi_{yy} \space d\Omega +
        \int_\Omega \overline{U} \space \phi_{zz} \space d\Omega ) 
        + (
        \int_\Omega \overline{u'^2} \space \phi_x \space d\Omega + 
        \int_\Omega \overline{u'v'} \space \phi_y \space d\Omega + 
        \int_\Omega\overline{u'w'} \space \phi_z \space d\Omega).
    \end{gathered}
\end{equation}
The resulting equation-space coordinates inherit the smoothing properties of the integral operator and remain well-defined for arbitrarily high derivative orders, since each additional integration by parts simply transfers another derivative onto the analytical test function. They also extend naturally to unstructured and curvilinear meshes due to numerical quadratures accommodating spatially varying cell sizes without modification, whereas finite-difference stencils require lossy upsampling to a structured grid and rapidly lose accuracy as the upsampling factor grows. Together, these properties make the weak formulation directly applicable to the regime in which equation-space analysis has historically failed---noisy, high-order, geometrically complex measurements. 

\subsection{A fully derivative-free equation space for incompressible flow} \label{sec: 2DIncompressibleRANS}

A single integration by parts is sufficient to remove all derivatives from linear terms in the governing PDE, but nonlinear terms generally retain at least one derivative on the data unless they can be expressed in conservation form. Here we show that, for two-dimensional incompressible flow, the advective nonlinearity admits such a restructuring, yielding an equation-space construction in which \emph{no numerical derivative of the data appears anywhere}. The same construction extends immediately to three dimensions and to any incompressible advective system.

\textbf{A conservation-form perspective.} Under two-dimensional incompressibility, the mean advective nonlinearity can be rewritten exactly in conservation form,
\begin{equation}
\label{eq: ConservationForm}
  \overline{U}\,\overline{U}_x + \overline{V}\,\overline{U}_y \;=\;
  \partial_x\!\ (\overline{U}^{\,2}) +
  \partial_y\!\left(\overline{U}\,\overline{V}\right),
\end{equation}
since $\overline{U}\,\overline{U}_x = \tfrac{1}{2}(\overline{U}^{\,2})_x$ and $\overline{V}\,\overline{U}_y + \overline{U}\,\overline{V}_y = (\overline{U}\,\overline{V})_y$ with $\overline{U}_x + \overline{V}_y = 0$ adding a vanishing term. Multiplying~\eqref{eq: ConservationForm} by a compactly supported test function $\phi$, integrating over $\Omega$, and applying integration by parts \emph{once} on each term offloads the remaining derivatives onto $\phi$:
\begin{equation}
\label{eq: AdvectionWeak}
  \int_\Omega \left(\overline{U}\,\overline{U}_x +
                    \overline{V}\,\overline{U}_y\right) \phi \,
  \mathrm{d}\Omega
  \;=\;
  -\!\int_\Omega \overline{U}^{\,2}\,\phi_x \,
  \mathrm{d}\Omega
  \;-\;
  \int_\Omega \overline{U}\,\overline{V}\,\phi_y\,\mathrm{d}\Omega.
\end{equation}
The full streamwise RANS equation in 2D therefore takes the fully derivative-free weak form
\begin{equation}
\label{eq: WeakRANS_2D_CompactForm}
\begin{gathered}
  -\!\int_\Omega \overline{U}^{\,2}\,\phi_x\,\mathrm{d}\Omega
   -\int_\Omega \overline{U}\,\overline{V}\,\phi_y\,\mathrm{d}\Omega
   \;=\;
   \int_\Omega \tfrac{\overline{p}}{\rho}\,\phi_x\,\mathrm{d}\Omega
  \\
   + \nu\!\left(\int_\Omega \overline{U}\,\phi_{xx}\,\mathrm{d}\Omega +
                \int_\Omega \overline{U}\,\phi_{yy}\,\mathrm{d}\Omega\right)
   + \int_\Omega \overline{u'^{\,2}}\,\phi_x\,\mathrm{d}\Omega
   + \int_\Omega \overline{u'v'}\,\phi_y\,\mathrm{d}\Omega.
\end{gathered}
\end{equation}
Every coordinate of the equation-space vector (Eq.~\eqref{eq:eqn_space}) is now an inner product of a measured field against an analytically known derivative of a test function. The manipulation generalizes immediately to three dimensions and to any incompressible advection nonlinearity that admits a conservation form.

\paragraph{A two-order derivative reduction for higher-order PDEs.}
For governing equations that do not admit a conservation-form, including the vorticity-transport equation analyzed in Sec.~\ref{sec:duct_vorticity}, rewriting the weak formulation still reduces the maximum derivative order applied to the data, often dramatically. The third-order dissipative term in the vorticity-transport equation, for instance, becomes
\begin{equation}
\nu \int_\Omega \frac{\partial^2 \Omega^x}{\partial n^2} \phi \, d\Omega
\;=\;
\nu \int_\Omega \Omega^x \phi_{nn}\, d\Omega
\label{eq:vorticity_reduction}
\end{equation}
after two integrations by parts, replacing a third-order derivative of
velocity with the first-order derivative required only to construct the vorticity itself. This two-order reduction is the structural reason that the first data-driven dominant-balance decomposition of a third-order PDE is made possible (Sec.~\ref{sec:duct_vorticity}).

\section{Results}

In this section, we are now ready to apply the weak dominant balance algorithm to several fluid flow problems, each embodying a unique set of challenging dynamics where gradients are not easily calculated. 
First, we demonstrate the noise robustness of the weak method by directly comparing it to a pointwise (finite difference) baseline on a noisy example of transitional flow to turbulence over a flat plate---a tweaked test case from~\cite{CallahamDB} with no-noise results that are available as a ground truth.  
Next we explore some canonically interesting and industrially relevant flows where pointwise dominant balance fails, and where it is necessary to use the weak form to learn the underlying dynamics-based clustering. Specifically, we consider the turbulent duct, which contains critical mechanisms in the third-order vorticity transport equations that generate corner vortices, which are structures that affect wall-shear and mean velocity contours. We also investigate turbulent flow through a wavy channel, which requires dominant balance to account for the non-rectilinear mesh over the curved geometry of the wall. For the wavy channel, both DNS and particle-image velocimetry (PIV) data at equivalent $Re_\tau$ are considered, and the corresponding datasets serve as the ideal test case to evaluate weak dominant balance's ability to handle noise in a real-life experimental setting.

\subsection{Noisy Flat Plate Turbulent Boundary layer}
\label{sec: TBL_NoiseStudy}
 
Our first benchmark quantifies the resilience of the weak equation-space construction under controlled additive measurement noise, using the canonical transitional flat-plate boundary layer as a rigorously characterized reference case~\cite{StreaksSpotsTransistion_Zaki2013, JHLTBD_2007, JHLTBD_2008}. This flow spans five physical regimes canonical to external aerodynamics---freestream, laminar, transitional, inertia-dominated, and viscous-dominated turbulent regions---each of which maps to a well-separated structure in the noise-free equation-space embedding~\cite{CallahamDB}, providing an unambiguous ground truth against which to measure the degradation of any equation-space construction under noise.

\begin{figure}
    \begin{center}
        \includegraphics[width=\textwidth]{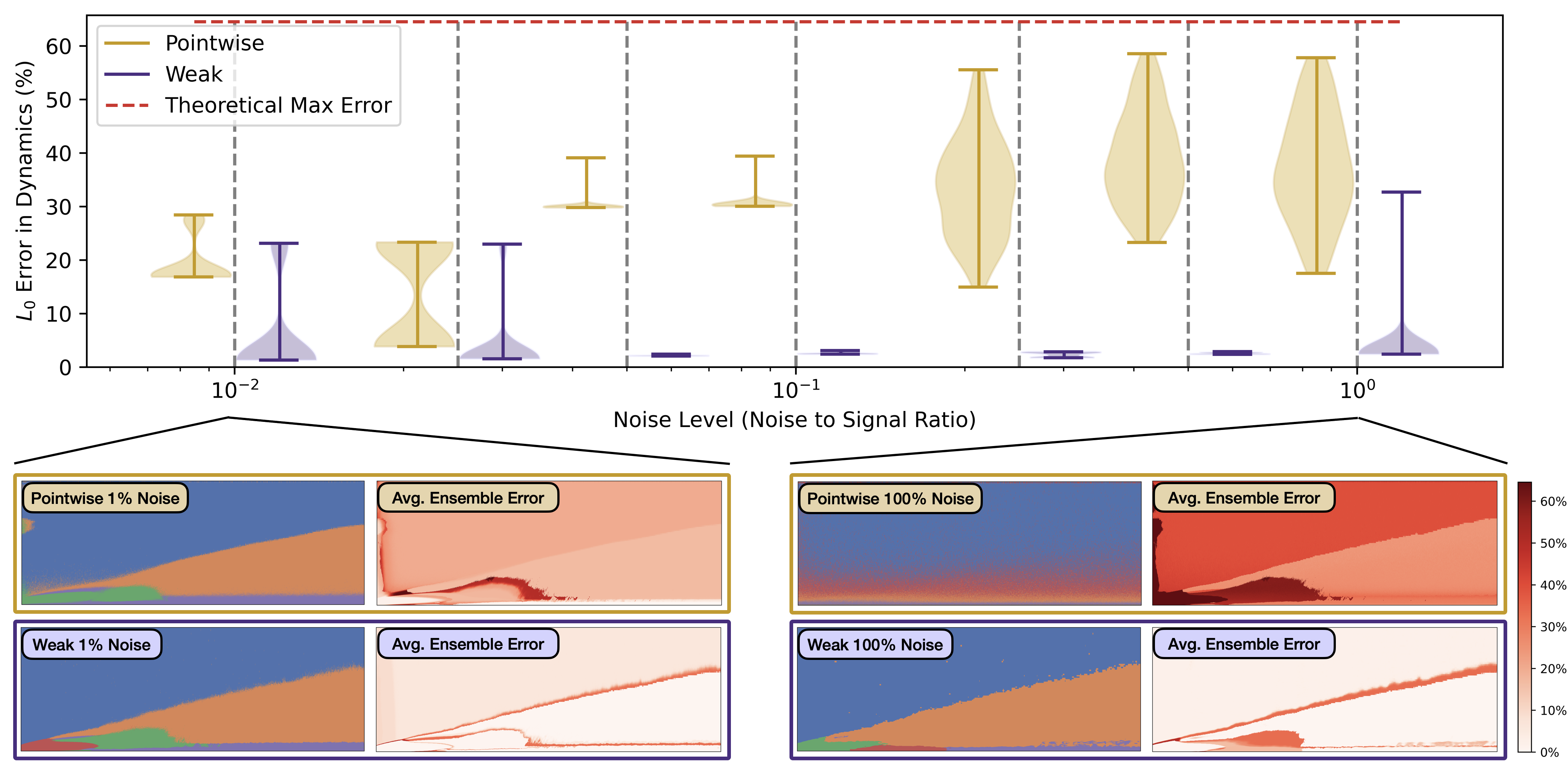}
    \end{center}
    \caption{\textbf{The weak formulation sustains accurate regime identification at noise levels two orders of magnitude beyond the breakdown point of pointwise differentiation.} \textbf{Top:} Distributions of $\mathcal{L}_0$ classification error (the fraction of spatial points whose identified active terms disagree with the noise-free ground truth) for the weak formulation (purple) and the pointwise (finite-difference) baseline (yellow), evaluated on the transitional flat-plate boundary layer of Fig.~\ref{Fig:Overview}. Each violin aggregates 100 independent trials in which zero-mean Gaussian noise of the specified noise-to-signal ratio was added independently to the mean fields. Whiskers mark the minimum, median, and maximum of each distribution; the dashed red line indicates the theoretical maximum error ($\sim$66\%), corresponding to the trivial identification of no dynamics across the entire domain. The weak formulation maintains a median error below 2.71\% across the full range from 1\% to 100\% noise, while the pointwise baseline degrades to a median error above $\sim$30\% already at 5\% noise. \textbf{Bottom:} Representative spatial decompositions at 1\% and 100\% noise. Left of each pair: the domain partition from a single representative trial. Right of each pair: the per-point ensemble-mean error across all 100 trials at that noise level (red colorbar, 0--60\%). At 1\% noise the pointwise baseline already fragments the physical regimes; at 100\% noise the weak decomposition remains qualitatively consistent with the noise-free reference (see Fig.~\ref{Fig:Overview} for the baseline breakdown of regions/dynamics absent any noise). The bimodality visible in the pointwise distributions at low noise reflects two near-degenerate clustering optima and is discussed in Sec.~\ref{sec: TBL_NoiseStudy}.}
    \label{Fig:2QuantComparison}
\end{figure}

Across all trials at each noise level (Fig.~\ref{Fig:2QuantComparison}), the weak construction sustains a median $\mathcal{L}_0$ classification error below 2.71\% even at noise-to-signal ratios of 100\%, while pointwise differentiation breaks down quantitatively and qualitatively at 1\% noise---a robustness gap of two orders of magnitude in the admissible noise level. At 1\% noise, the example decompositions under pointwise differentiation already lose physical meaning, with the merging of the laminar and transitional regions, whereas the weak decomposition remains indistinguishable from the noise-free reference up to 50\% noise and recovers four out of the the five canonical regimes through 100\% noise.

The error distributions also expose a sharp asymmetry between the repeatability of the two equation-space constructions. Under pointwise differentiation, the clustering objective develops two (near-degenerate) local minima at 1\% and 2.5\% noise, producing a bimodal error distribution and forcing case-by-case selection of the sPCA sparsity threshold and GMM component count to recover physical regimes. The weak construction shows no such fragility: the integral smoothing inherent to the formulation stabilizes the equation-space embedding into a single, well-separated manifold across the entire noise range, and a single set of hyperparameters recover the reference decomposition. The hyperparameter sensitivity of the pointwise baseline is therefore not a property of dominant-balance analysis as such, but a consequence of differentiating noisy fields and well-addressed by moving to a weak construction. Promising directions for fully automated hyperparameter selection ~\cite{tran2025weakformscientificmachine, KAISER2022105496} are discussed in Sec.~\ref{Sec: Discussion}.

\textbf{Error metric and label alignment.} We perform a direct pointwise comparison between the labeled \emph{dynamics} identified in the no-noise baseline and each noisy trial, which is representative of both steps in data-driven dominant balance---GMM followed by sPCA reduction. For the given RANS equation space, each point in the spatial domain has 6 terms that can be identified as either active or inactive. The sPCA reduction step produces a grid of booleans to signify active or inactive, with size $(no. \ of \ spatial \ grid \ points \times 6)$, and the reported $\mathcal{L}_0$ error per trial is the fraction of labels across the spatial domain that disagree. To verify that the reported trends are not artifacts of the metric itself, we have repeated the entire analysis using the adjusted Rand index~\cite{hubert1985comparing} and the normalized mutual information~\cite{vinh2010nmi}, both of which are invariant to label permutation by construction and weight cluster-boundary disagreements differently from $\mathcal{L}_0$. All three metrics yield the same ordering of methods at every noise level (see~\ref{sec: Appendix_ErrorMetric}), and the qualitative conclusion---weak dominant balance sustains low error through $100\%$ noise while the pointwise method degrades at $1\%$---is independent of metric choice. We retain $\mathcal{L}_0$ as the primary reported quantity because its interpretation (``fraction of point-wise misclassified dynamics'') is the most directly physical.

\subsection{Turbulent Duct Flow} \label{sec:duct_vorticity}

Having established quantitative robustness of the weak method, we now turn to a system that was previously inaccessible to dominant balance analysis---turbulent flow through a square duct. From a physical point of view, the mechanistic origin of corner vortices that warp streamwise-velocity contours along the bisector and redistribute wall shear in the corner region is a long standing question of turbulent duct flow~\cite{Vinuesa_OGDuctARPaper_2014, Vinuesa2016ConvergenceOfNumSimsOfWallBoundedAndDuctFlows}. The RANS perspective can be used to capture the impact of secondary motions of the second kind (corner vortices), appearing there as a persistent cross-stream advection, but the cause of these motions only appears as a balance between Reynolds-stress gradients in the vorticity transpose equations. Resolving this balance requires an equation-space analysis at third derivative order, a regime in which pointwise differentiation amplifies measurement noise so severely that the embedding loses its manifold structure even on noise-free DNS data (Fig.~\ref{fig:WeakVanillaDuctResults}). By comparing the RANS and vorticity transport decompositions, we recover that, indeed, the mechanisms underlying Prandtl's secondary motions are more clearly encoded by the vorticity transport perspective. 
This asserts the broader implication: fundamental mechanisms that are only captured by higher-order equations---including physical systems beyond vortically dominated flows---are now within scope of dominant balance analysis because of our weak formulation. 

\subsubsection{Second-Order Equation Space (Streamwise RANS Equation)}

\begin{figure}[t!]
    \begin{center}
        \includegraphics[width=\textwidth]{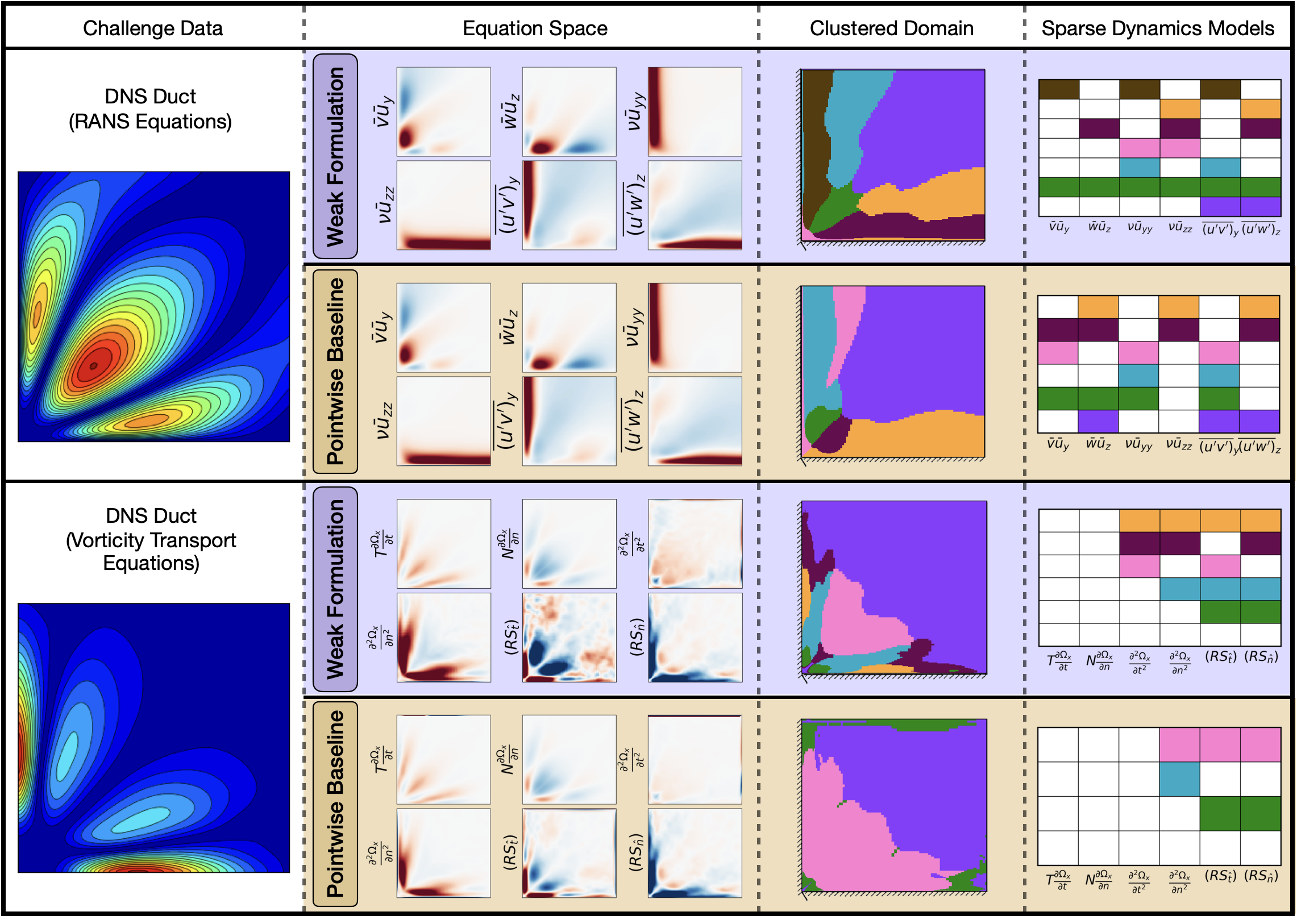}
    \end{center}
    \caption{\textbf{Weak dominant balance is more accurate in low-order PDEs and generalizes to high-order PDEs.} Each row presents a complete equation-space analysis of the duct, with a side by side comparison of the weak formulation against a pointwise (finite difference) baseline, and we have used the same data, the same governing equations, and the same clustering pipeline (GMM + sPCA) to isolate the differences between approaches. Columns from left to right: (i) a representative field of the input data used for both approaches (mean streamwise velocity for RANS analyses, mean streamwise vorticity $\Omega^x$ for the vorticity transport analysis); (ii) the per-term weak equation-space coordinates, each shown over the analysis domain; (iii) the unsupervised cluster labels obtained by Gaussian-mixture modelling of points in equation space, projected back onto the spatial domain; and (iv) the sparse-PCA dynamics matrix identifying which equation-space terms are active (colored) or negligible (white) in each cluster. In the case where no terms are identified as dominant, the color of the inactive region defaults to purple in the clustered domain and the bottom row in the sparse dynamics matrix is left white. \textbf{Row 1:} Turbulent square duct under the streamwise Reynolds-averaged Navier--Stokes equation (bottom-left quadrant shown). The decomposition captures the symmetry across the corner bisector, the wall and corner shear layers, and the persistent spanwise advection induced by the corner vortices. \textbf{Row 2:} The overall structure is approximately captured, as this remains a second-order PDE on a rectilinear grid, but the symmetry is no longer present and sPCA identifies physically implausible active terms, including several misclassified spanwise advection contributions, which are the motivating effects of this canonical flow. \textbf{Row 3:} The same duct flow analyzed under the third-order streamwise vorticity-transport equation---to our knowledge, the first data-driven dominant-balance decomposition of a third-order PDE. The weak formulation reduces the maximum derivative order acting on the data from three to one (Eq.~\ref{eq:vorticity_reduction}), isolating the deviatoric Reynolds shear stress $RS_{\hat t}$ and the cross-stream Reynolds-stress difference $RS_{\hat n}$ as the mechanisms responsible for corner-vortex generation. \textbf{Row 4:} The third order derivatives in the dissipation term amplify finite-difference truncation error to the point that the equation-space embedding loses it's manifold structure. The resulting partition collapses into  single dominant cluster covering most of the domain that bears little relation to the known mechanism of corner-vortex generation, extending beyond the reasonable reach of these phenomena.
    }
\label{fig:WeakVanillaDuctResults}
\end{figure}

The first row of Fig.~\ref{fig:WeakVanillaDuctResults} from left to right shows the reconstructed equation space, domain decompositions, and corresponding sparse dynamics models under the streamwise RANS equation for the bottom left quadrant of turbulent duct flow. Immediately, the symmetry in the domain decomposition across the bisector, the concavity of the flow structure at the corner, and the lifting at the duct centerlines (the border between each quadrant) reflect known structures in the duct~\cite{Vinuesa_OGDuctARPaper_2014} that are not explicitly enforced in our method, but captured by weak dominant balance. The results of sPCA show high shear regions identified on both walls and on the either border of the duct core (light-blue and orange regions in Fig.~\ref{fig:WeakVanillaDuctResults}), and these dynamics are consistent with those expected from wall effects alone. In addition to the viscous and Reynolds stress terms, the spanwise advection terms are also identified as active in regions along both walls (brown and burgundy) and in a pocket nestled in the corner itself (green). This added momentum transfer in the spanwise directions is not present in typical boundary layers or channel flows. It is only identified here due to a persistent spanwise velocity that is introduced by the corner vortices in the duct flow, and these challenging effects of momentum transport are successfully captured by weak dominant balance.

\subsubsection{Third-order Equation Space (Streamwise Vorticity Transport Equation)}

While the effects of secondary motions are present in the momentum equation, the mechanisms that cause these motions are only identifiable in the streamwise vorticity transport equations, which can be written:
\begin{equation}
\label{eqn: FullVortTrans}
    \begin{gathered}
    \overline{U} \frac{\partial \Omega^x}{\partial x} + \overline{V} \frac{\partial \Omega^x}{\partial y} + \overline{W} \frac{\partial \Omega^x}{\partial z} = \nu (\frac{\partial^2 \Omega^x}{\partial x^2} + \frac{\partial^2 \Omega^x}{\partial y^2} + \frac{\partial^2 \Omega^x}{\partial z^2}) + \\ \Omega^x\frac{\partial \overline{U}}{\partial x} + \Omega^y\frac{\partial \overline{U}}{\partial y} + \Omega^z \frac{\partial \overline{U}}{\partial z} + (\frac{\partial^2}{\partial y^2} - \frac{\partial^2}{\partial z^2})(\overline{-vw})+ \frac{\partial^2}{\partial y \partial z}(\overline{v^2} - \overline{w^2}),
    \end{gathered}
\end{equation}
derived by taking the curl of the RANS equation to describe the transport of $x,y,z$ mean vorticity fields $\Omega^x, \Omega^y, \Omega^z$. Statistical homogeneity in the streamwise direction implies $\partial_x \langle\,\cdot\,\rangle = 0$ for any mean quantity in the fully developed duct, eliminating the streamwise-gradient terms in Eq.~\eqref{eqn: FullVortTrans} and reducing it to:
\begin{equation}
    \label{eqn: VortTransEuclidean_NoZGradients}
    \begin{gathered}
    \overline{V} \frac{\partial \Omega^x}{\partial y} + \overline{W} \frac{\partial \Omega^x}{\partial z} = \nu (\frac{\partial^2 \Omega^x}{\partial y^2} + \frac{\partial^2 \Omega^x}{\partial z^2}) +
    \\
    \Omega^y\frac{\partial \overline{U}}{\partial y} + \Omega^z \frac{\partial \overline{U}}{\partial z} + (\frac{\partial^2}{\partial y^2} - \frac{\partial^2}{\partial z^2})(\overline{-vw})+ \frac{\partial^2}{\partial y \partial z}(\overline{v^2} - \overline{w^2}).
\end{gathered}
\end{equation}
In order to preserve the direction of vorticity across the bisector, the data is cast into normal/tangential coordinates ($t, n$ for tangential and normal directions and $T,N$ for tangential velocity and normal velocity). Also note that the Reynolds stress terms from the full equation are condensed into two terms, one containing the deviatoric Reynolds shear stress, $\overline{vw}$, and the other containing the cross stream Reynolds stress difference, $\overline{v^2} - \overline{w^2}$. These terms (condensed into $RS_{\hat{t}}$ and $RS_{\hat{n}}$ respectively in Fig.~\ref{fig:WeakVanillaDuctResults} and in Eq.~\eqref{eqn: VortTransNormTan}) are directly associated with producing secondary motions of the second kind, and will be the focus of the decomposition. Note that secondary motions of the first kind are not present in the duct, and thus, the vortex stretching terms associated with this phenomena in Eq.~\eqref{eqn: VortTransEuclidean_NoZGradients} vanish to zero. The final simplified form of our vorticity transport equation in normal/tangential coordinates can now be written:
\begin{gather}
\label{eqn: VortTransNormTan}
    T \frac{\partial \Omega^x}{\partial t} + N \frac{\partial \Omega^x}{\partial n} = \nu \frac{\partial^2 \Omega^x}{\partial t^2} + \nu\frac{\partial^2 \Omega^x}{\partial n^2} + RS_{\hat{t}} + RS_{\hat{n}}.
\end{gather}
The corresponding weak coordinate for the wall-normal dissipative term, after two integrations by parts and use of the compact support of $\phi$, reads
\begin{equation}
\label{eq: WeakViscousVorticity}
  \nu \int_\Omega \frac{\partial^{\,2}\Omega^x}{\partial n^{\,2}}\,\phi\,
       \mathrm{d}\Omega
  \;=\;
  \nu \int_\Omega \Omega^x\,\phi_{nn}\,\mathrm{d}\Omega,
\end{equation}
which replaces a third-order derivative of the velocity field in $\frac{\partial^{\,2}\Omega^x}{\partial n^{\,2}}$ with only the first-order derivatives of velocity required to construct $\Omega^x$ against an analytically derived $\phi_{nn}$. This two-order reduction in the derivative applied to the data is the reason that weak dominant balance succeeds on the vorticity transport equation whereas the noise amplification of repeated finite-differencing causes the pointwise reconstruction to fail (Fig.~\ref{fig:WeakVanillaDuctResults}, fourth row). 

To our knowledge, Fig.~\ref{fig:WeakVanillaDuctResults} (third row) is the first data-driven dominant-balance decomposition of a third-order PDE. The pointwise baseline cannot produce this decomposition (fourth row) since the third-order derivatives in the dissipative terms of Eq.~\eqref{eqn: VortTransNormTan} amplify finite-difference noise so severely that the equation-space embedding loses its manifold structure and the GMM clusters become physically meaningless. By reducing the maximum derivative order acting on the data from three to one (Eq.~\eqref{eq: WeakViscousVorticity}), the weak formulation restores the embedding, yielding a decomposition that recovers the canonical mechanism of secondary-flow generation in a turbulent duct without supervision. This includes a corner band in which both the deviatoric Reynolds shear stress $RS_{\hat{t}}$ and the cross-stream Reynolds-stress difference $RS_{\hat{n}}$ are simultaneously active, an extended region around the bisector in which $RS_{\hat{t}}$ dominates, and pockets near either wall of strong balance between these turbulent effects and the vortical dissipation terms.

The area of the domain covered by these regions matches the high-probability vortex-occurrence map at the same $Re_\tau \approx 180$ reported by~\cite{Vinuesa2016ConvergenceOfNumSimsOfWallBoundedAndDuctFlows}. The decomposition explains both of the well-known mean-flow signatures of Prandtl's secondary motions of the second-kind---the warping of streamwise velocity contours along the bisector and the corner-localized redistribution of wall shear---as expressions of the dominant terms identified by weak dominant balance. This overlap between our identified regions and the structures of prior analyses additionally confirms that the critical physics of the turbulent duct are better encoded by the approach using the vorticity transport equations, an embedding that is only made possible by our weak formulation.

\subsection{Bridging Simulation and Experiment: Turbulent Wavy Channel Flow}
\label{Sec:WavyWallResults}

Whether an equation-space decomposition reflects genuine physics or numerical artifacts of one particular dataset can be tested directly when independent observations of the same flow are available. We analyze a turbulent wavy channel using a matched direct numerical simulation (DNS) and an independent particle-image-velocimetry (PIV) experiment at the same friction Reynolds number, $Re_\tau \approx 206$
\cite{Lagemann_WavyWall_OpticalFlow_2024,WavyWallDataGen_PhysRevFluids.4.034605,lagemann2024impact}. Despite substantial differences in spatial resolution, noise statistics, and missing-data patterns between the two datasets, the weak formulation recovers nearly identical decompositions from both.

\begin{figure}[t!]
    \begin{center}
        \includegraphics[width=\textwidth]{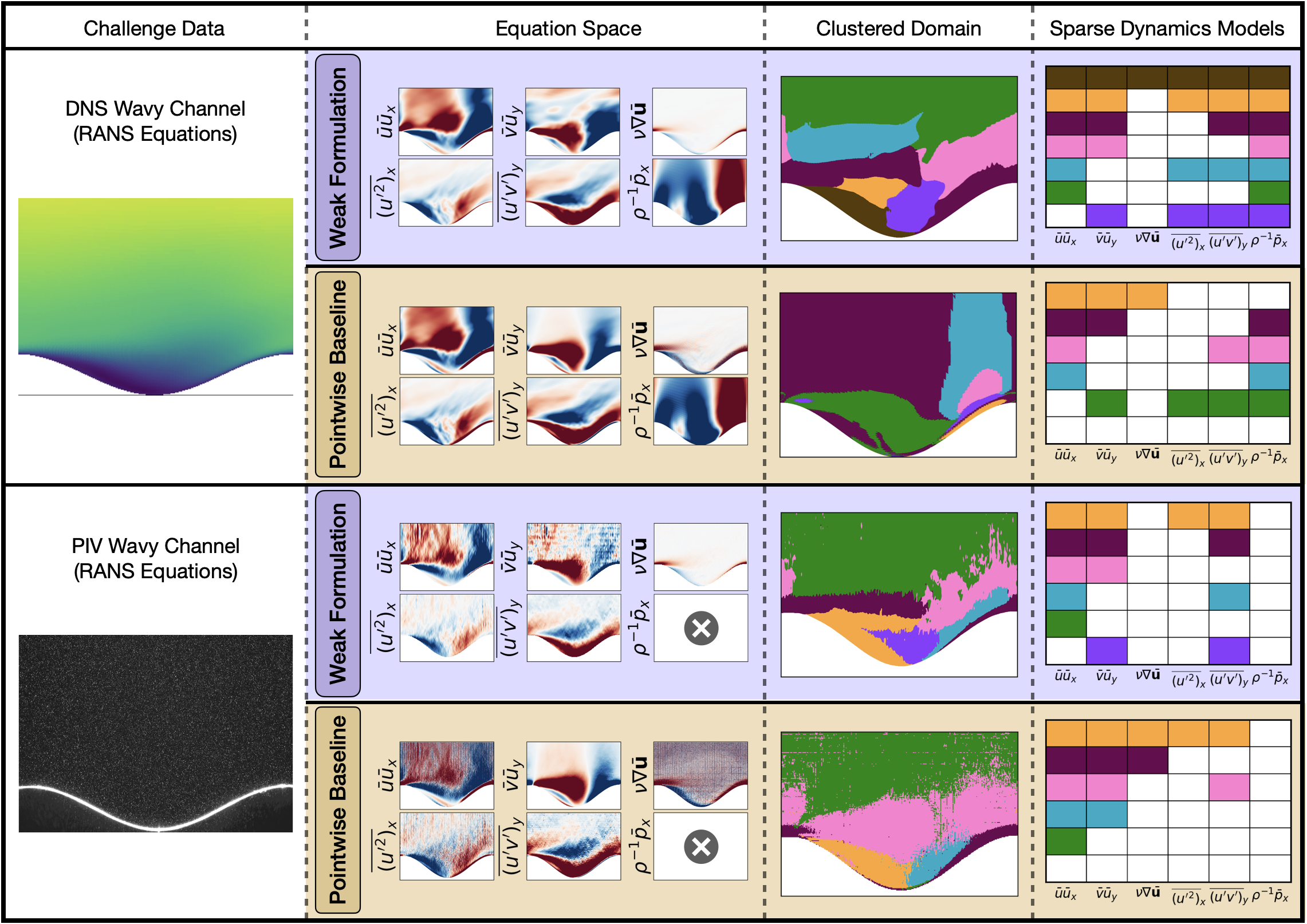}
    \end{center}
    \caption{\textbf{Weak dominant balance generalizes to real-world experimental measurements and non-rectilinear geometries.} Rows and columns partition results the same as in Fig.~\ref{fig:WeakVanillaDuctResults}, the representative fields in column (i) now being mean streamwise velocity for DNS analyses and raw particle-image-velocimetry snapshot for the experimental case, in column (ii), rows 3 and 4, empty $\rho^{-1}\bar{p}_x$ fields reflect the lack of pressure measurements available from PIV, and column (iv) includes an empty pressure column for the sake of matching the DNS dynamics matrix. \textbf{Row 1:} Turbulent flow over a wavy channel analyzed by direct numerical simulation under the 2D incompressible RANS equation. The non-rectilinear mesh, handled directly by numerical quadrature, presents no obstacle to the weak construction. \textbf{Row 2:} Lossy upsampling onto a structured grid (require to appl finite-difference stencils to the curvilinear mesh) prduces several misclassifications, including the freestream and recirculation regions being assigned to the same cluster despite opposing dynamics \textbf{Row 3:} The same wavy channel analyzed from an independent particle-image-velocimetry experiment at matched friction Reynolds number ($Re_\tau \approx 206$). The decomposition agrees with the DNS row in both spatial structure and identified dynamics (Jaccard $J_{\text{weak}} = 0.524$), despite the substantial noise and missing-data patterns characteristic of PIV; both rows additionally resolve a compact regime in the wave trough (purple) that has not previously been identified as dynamically distinct. Cluster colors are matched across rows 3--4 by Hungarian assignment. \textbf{Row 4:} Measurement noise from the experimental data propogates through the differentiation stencils into the equation-space coordinates, blurring cluster boundaries and supressing turbulent contributions in the near-wall region.}
\label{fig:WeakVanillaWavyChannelResults}
\end{figure}

In Fig.~\ref{fig:WeakVanillaWavyChannelResults}, both PIV and DNS decompositions manage to identify the freestream (green), shear layer (burgundy), and recirculation (brown in DNS, orange in PIV) using the weak approach,  which all closely match in structure and in identified dynamics between datasets. Notably, these major features match the reattachment pattern documented in~\cite{WavyWallDataGen_PhysRevFluids.4.034605}, the mean streamwise velocity contours reported in~\cite{YOON2009697}, and the dominance of Reynolds shear stress throughout the trough in~\cite{Hudson:1996TurbulenceProduction}. However, the weak decomposition offers a more direct delineation of distinct regimes compared previous analyses, differentiating between the balances that are specific to phenomena of shear separation, detachment in the recirculation region, or acceleration along the up-slope of the wave.

\textbf{Quantitative DNS--PIV agreement.} Despite the noisy, incomplete data produced by the nature of PIV, we find that across the cluster assignments shown in Fig.~\ref{fig:WeakVanillaWavyChannelResults}, weak dominant balance achieves a Jaccard index of $J_{\text{weak}} = 0.5240$ between the DNS and PIV decompositions, compared to $J_{\text{orig}} = 0.3763$ when using finite difference (cluster labels matched by Hungarian assignment to maximize overlap). The corresponding adjusted Rand indices are $\text{ARI}_{\text{weak}} = 0.4497$ and $\text{ARI}_{\text{orig}} = 0.2241$. These metrics quantitatively confirm the qualitative findings in Fig.~\ref{fig:WeakVanillaWavyChannelResults}, that the weak formulation produces decompositions that transfer between simulation and experiment with minimal loss, while the pointwise method does not. 

Interestingly, both the DNS and PIV decompositions resolve a compact region in the trough (purple in Fig.~\ref{fig:WeakVanillaWavyChannelResults}) whose dominant balance differs from the surrounding recirculation, shear-layer, and freestream regimes. The active terms identified by the sparse dynamics model are the wall-normal mean advection and the wall-normal Reynolds-stress gradient $\overline{u'v'}_y$, a combination consistent with streamlines bending toward the wall as the recirculation bubble closes and momentum is redistributed across the upper edge of the trough. The region is spatially bounded upstream by the recirculation bubble, downstream by the accelerating up-slope flow, and coincides with the documented sign change in the streamwise pressure gradient~\cite{WavyWallDataGen_PhysRevFluids.4.034605}---placing it physically at the transition between detachment and reattachment. To our knowledge, this regime has not previously been isolated as dynamically distinct, either by the classical investigations of the wavy channel~\cite{Hudson:1996TurbulenceProduction, WavyWallDataGen_PhysRevFluids.4.034605, YOON2009697} or by pointwise equation-space analysis as seen clearly in Fig. \ref{fig:WeakVanillaWavyChannelResults}.

Three independent checks support a physical, rather than algorithmic, origin: (i) the regime appears in both the DNS and the PIV decompositions, which carry very different noise statistics; (ii) it persists under perturbation of the test-function support radius across the full range used elsewhere in this work; and (iii) it survives random reinitialisation of the GMM seed and small variations in the sPCA sparsity threshold. It is, we believe, a concrete example of the kind of regime resolution that becomes accessible once equation-space dominant-balance analysis is no longer constrained by explicit numerical differentiation.

\subsection{Wall Clock Speed of Equation Space Construction}
\label{sec:FFTSpeedDemon}

\begin{table}[t!]
\caption{\textbf{Weak dominant balance scales seamlessly to large computational grids.} The speed of equation space computation for the weak and finite difference approaches is compared across the variations in domain size and equation spaces, which are separated into columns. The Domain Size row reports the number of points on which gradients are evaluated, and the reported Wall Times compare the number of seconds it takes to complete either the weak or pointwise evaluation of terms in an equation space. The code is optimized in JAX to run on a GeForce RTX 2080 Ti GPU with 11GB of dedicated memory, and no other task was running on the machine while generating these benchmarks. The code is CPU compatible, although generally with slow downs in run time.}
\centering
\label{tab:example}

\begin{tabular}{
    l
    c
    c
    c
    c
    c
}
\toprule
Algorithm & \makecell{Turbulent Duct \\ (RANS)} & \makecell{Turbulent Duct \\ (Vorticity Transport)} & \makecell{Wavy Wall \\ DNS (RANS)} & \makecell{Wavy Wall \\ PIV (RANS)} \\
\midrule

\makecell{Domain Size\\ (\# of grid points)}
& 9409
& 9409
& 1,390,217
& 3,090,960

\\[2ex]

\makecell{Pointwise Wall Time\\ (seconds)}
& 0.96
& 1.09
& 2262.16
& 9959.68 

\\[2ex]

\makecell{Weak Wall Time\\ (seconds)}
& 1.16
& 2.48
& 3.53
& 12.89

\\
\bottomrule
\end{tabular}

\end{table}

In addition to deficiencies in accuracy, pointwise finite differencing suffers poor computational scalability to high-dimensional domains. Without proper sparsification, finite difference stencils quickly exceed memory limits for systems that span more than a single dimension. In our case, however, the severe additional run time that scales poorly with the size of the system is a major bottle neck for a dominant balance practitioner. In Table~\ref{tab:example}, we compare the wall clock time it takes to apply pointwise and weak-form differentiation to construct equation-space fields in each of our four cases. While the additional overhead of the weak implementation causes slightly slower evaluations on the small duct grid, the method scales seamlessly to significantly larger spatial domains by evaluating the convolution of data and test functions as pointwise multiplication in Fourier space. Finite-difference instead requires repeated, costly multiplications of large matrices to evaluate spatial gradients, leading to $\sim 700 \times$ speedup by weak dominant balance for our largest system. 

\section{Discussion} \label{Sec: Discussion}

This work has introduced \emph{weak dominant balance}, an approach to enabling dominant balance analysis on high-dimensional, multi-scale physical systems. The method incorporates recent advances in scientific machine learning with a classical, equation-based approach to address several levels of complicated, previously inaccessible data. Specifically, we show that our framework is essential to handle noisy data, higher-order derivatives, and complex geometries. The immediate benefits are realized in a rigorous, quantitative evaluation on the flat plate boundary layer where weak dominant balance delivers more accurate decompositions at $100\%$ noise than a comparable pointwise baseline at $1\%$ noise. The method is then tested on several turbulent flows with various additional challenges, where no known dominant balance scalings currently exist. 

Analysis of the turbulent duct under streamwise RANS captures the impact of secondary motions, highlighting the added streamwise advection which warps velocity contours and wall shear distributions. Under the third-order streamwise vorticity transport equation, regions dominated by turbulent effects (the mechanisms responsible for persistent secondary motions in the duct) are identified, the extent of which closely matches prior probabilistic characterizations of the location of corner vortex events. This improved representation of the corner vortex regions under the vorticity transport equation demonstrates that a weak embedding facilitates a more accurate representation of entirely new classes of physical mechanisms---including additional vortically dominated flows or any other application that may be described by higher-order PDEs. 

We finally extend our analysis to the wavy channel flow, extracting regions that correspond to the many instances of canonical physics in this example. 
The non-rectilinear mesh of the data, which is traditionally challenging for explicit numerical differentiation, did not prevent the weak form from identifying a decomposition that aligns with structures identified in previous investigations of the flow. Weak dominant balance was also able to overcome the experimental challenges of PIV data from a wavy channel that was the physical twin to our DNS data. Not only was the resulting decomposition from experimental data nearly identical to the DNS companion, but the relevant underlying dynamics also match closely in each region while only using fields acquired through real-world measurements. The results of our testing show the marked ability of the weak approach in both structure and dynamics identification for challenging examples of turbulent flows.

\paragraph{Broader applicability.}
Although our demonstrations are drawn from fluid mechanics, the construction is physics-agnostic. Any spatiotemporal system with a known candidate governing PDE and accessible state observations is amenable to the same weak equation-space analysis. Within fluid dynamics, mixture-of-experts approaches to RANS and LES closure modelling~\cite{Lozano_Duran_Bae_2023, 2024_BuildingBlockSGSModel, Matai2019_ZonalEddyViscosity, BUCHANAN2026106899, AcceleratingSolvers_Otmani} rely on a regime-identification step that either requires immense amounts of training data for a neural-net classifier, or relies explicitly on dominant-balance based domain decompositions whose limitations---noisy measurements, high-order PDEs, irregular geometries---are precisely those addressed in this work~\cite{Roques2024_DBforRANS}. The same regime-identification capability extends to closed-loop control: reinforcement-learning environments for flow control such as HydroGym \cite{lagemann2025hydrogym_a, lagemann2025hydrogym_b} require compact, interpretable state abstractions, and a derivative-free, noise-robust equation-space partition offers a principled route to model-based control informed by the locally dominant physics rather than by black-box features alone.
Beyond fluids, candidate targets include reaction--diffusion dynamics in cardiac biology~\cite{Quarteroni_Manzoni_Vergara_2017, quarteroni2017integrated}, resistive magnetohydrodynamics in fusion~\cite{park2025kinetic, Kaptanoglu:2020vu}, several outlooks on the sparse multi-mechanism balances in geophysical systems~\cite{bodnar2025foundation, Vallis2016_GFDPerspective, SALCEDOSANZ2020256, Sonnewald_2021}, and the varying reactive--convective--diffusive processes active across a flame front~\cite{zeng2020complex}. In each of these settings, the obstacles to equation-space analysis we have addressed are not exceptional but the default. By eliminating the differentiation bottleneck that has confined equation-space analysis to idealized simulations, the weak formulation makes mechanism-level interpretation of governing PDEs available for the kinds of data that characterize most physical systems of practical interest.

\paragraph{Limitations and outlook.}
The weak formulation introduces the selection of the underlying basis function and its support as two principal hyperparameters whose joint selection currently relies on grid search guided by the physical length scales of interest. The downstream clustering pipeline requires the GMM component count and the sPCA sparsity threshold as two further hyperparameters. At very low noise levels we observe a residual ambiguity between adjacent GMM minima (Sec.~\ref{sec: TBL_NoiseStudy}), reflecting the absence of an unsupervised criterion for assessing physical fidelity of a candidate clustering. Recent developments in data-adaptive test-function recovery~\cite{tran2025weakformscientificmachine}, in alternative compactly supported bases~\cite{Bortz2023_WENDY_BumpFunctions}, and in automated cluster-number selection for dominant-balance analysis~\cite{KAISER2022105496} together offer a natural path to fully unsupervised operation. The structure of the equation-space manifolds themselves is an additional further direction. Nonlinear embedding techniques such as t-SNE~\cite{tSNE_vandermaaten_JMLR} or UMAP~\cite{UMAP_McInnes2018} may resolve entangled manifolds that are conflated by GMM or linear sPCA, and semi-supervised extensions could incorporate prior physical knowledge where available. We echo the sentiments across dynamics-based clustering literature, that the purpose of equation-space dominant-balance analysis has always been to augment, not supplant, physical expertise, and the developments above would broaden its applicability while preserving that role.

\section*{Code and Data Availability}
Data and code, including relevant configurations to generate each example in our results, can be found through our open-source GitHub repository \url{https://github.com/SamAhnert/Weak-Dominant-Balance.git}.

\section*{Acknowledgments}
We would like to acknowledge funding support from The Boeing Company and the National Science Foundation AI Institute in Dynamic Systems grant number 2112085. 
We also thank Nicholas Zolman for help reviewing and iterating on the method, and Nicholas Zolman, Sajeda Mokbel, and Ana Larrañaga-Janeiro for feedback on figures. 

\newpage

\bibliographystyle{plain}
 \begin{spacing}{.9}
 \small{
 \setlength{\bibsep}{4.8pt}
 \bibliography{references}

@article{Strommel_OceanCurrentIntensification,
    author = {Stommel, H.},
    title = {The westward intensification of wind-driven ocean currents},
    journal = {Eos, Transactions American Geophysical Union},
    volume = {29},
    number = {2},
    pages = {202-206},
    year = {1948}
}

@article{vinh2010nmi,
  author  = {Vinh, N. X. and Epps, J. and Bailey, J.},
  title   = {Information Theoretic Measures for Clusterings Comparison: Variants, Properties, Normalization and Correction for Chance},
  journal = {Journal of Machine Learning Research},
  year    = {2010},
  volume  = {11},
  number  = {95},
  pages   = {2837--2854},
}

@article{hubert1985comparing,
	author = {Hubert, L. and Arabie, P.},
	journal = {Journal of Classification},
	number = {1},
	pages = {193--218},
	title = {Comparing partitions},
	volume = {2},
	year = {1985}}

@article{AcceleratingSolvers_Otmani,
	author = {Otmani, K. E. and Mateo-Gabin, A. and Rubio, G. and Ferrer, E.},
	journal = {Engineering with Computers},
	number = {2},
	pages = {949--964},
	title = {Accelerating high order discontinuous Galerkin solvers through a clustering-based viscous/turbulent-inviscid domain decomposition},
	volume = {41},
	year = {2025}
}

@article{BUCHANAN2026106899,
	author = {Buchanan, T. and L{\u a}c{\u a}tu{\c s}, M. and West, A. and Dwight, R. P.},
	journal = {Computers \& Fluids},
	pages = {106899},
	title = {Data-driven RANS closures using a relative importance term analysis based classifier for 2D and 3D separated flows},
	volume = {305},
	year = {2026}}

@article{KAISER2022105496,
	author = {Kaiser, B. E. and Saenz, J. A. and Sonnewald, M. and Livescu, D.},
	journal = {Engineering Applications of Artificial Intelligence},
	pages = {105496},
	title = {Automated identification of dominant physical processes},
	volume = {116},
	year = {2022}}

@article{YOON2009697,
	author = {Yoon, H. S. and El-Samni, O. A. and Huynh, A. T. and Chun, H. H. and Kim, H. J. and Pham, A. H. and Park, I. R.},
	journal = {Ocean Engineering},
	number = {9},
	pages = {697-707},
	title = {Effect of wave amplitude on turbulent flow in a wavy channel by direct numerical simulation},
	volume = {36},
	year = {2009}}

@article{vinuesa2026explainableailearninglearners,
      title={Explainable AI: Learning from the Learners}, 
      author={Vinuesa, R. and Brunton, S. L. and Mengaldo, G.},
      year={2026},
      journal={arXiv preprint arXiv:2601.05525},
}

@article{ArranzLozanoDuran_IND_For_Turbulent_Flows, 
    title={Informative and non-informative decomposition of turbulent flow fields}, 
    volume={1000}, 
    journal={Journal of Fluid Mechanics}, 
    author={Arranz, G. and Lozano-Durán, A.}, 
    year={2024}, 
    pages={A95}
}

@article{OceanicDynamicsDB_2019,
    author = {Sonnewald, M. and Wunsch, C. and Heimbach, P.},
    title = {Unsupervised Learning Reveals Geography of Global Ocean Dynamical Regions},
    journal = {Earth and Space Science},
    volume = {6},
    number = {5},
    pages = {784-794},
    year = {2019}
}

@article{Global_Heating_2021,
    author = {Sonnewald, M. and Lguensat, R.},
    title = {Revealing the Impact of Global Heating on North Atlantic Circulation Using Transparent Machine Learning},
    journal = {Journal of Advances in Modeling Earth Systems},
    volume = {13},
    number = {8},
    year = {2021}
}

@article{EcologicallySimilarProvinces_2020,
	author = {Sonnewald, M. and Dutkiewicz, S. and Hill, C. and Forget, G.},
	journal = {Science Advances},
	number = {22},
	pages = {eaay4740},
	title = {Elucidating ecological complexity: Unsupervised learning determines global marine eco-provinces},
	volume = {6},
	year = {2020},
}

@article{SALCEDOSANZ2020256,
	author = {Salcedo-Sanz, S. and Ghamisi, P. and Piles, M. and Werner, M. and Cuadra, L. and Moreno-Mart{\'\i}nez, A. and Izquierdo-Verdiguier, E. and Mu{\~n}oz-Mar{\'\i}, J. and Mosavi, A. and Camps-Valls, G.},
	journal = {Information Fusion},
	pages = {256-272},
	title = {Machine learning information fusion in Earth observation: A comprehensive review of methods, applications and data sources},
	volume = {63},
	year = {2020},
}

@article{Vallis2016_GFDPerspective,
	author = {Vallis, G. K.},
	journal = {Proceedings of the Royal Society A: Mathematical, Physical and Engineering Sciences},
	number = {2192},
	pages = {20160140},
	title = {Geophysical fluid dynamics: whence, whither and why?},
	volume = {472},
	year = {2016},
}

@article{Sonnewald_2021,
	author = {Sonnewald, M. and Lguensat, R. and Jones, D. C. and Dueben, P. D. and Brajard, J. and Balaji, V.},
	journal = {Environmental Research Letters},
	number = {7},
	pages = {073008},
	title = {Bridging observations, theory and numerical simulation of the ocean using machine learning},
	volume = {16},
	year = {2021},
}

@article{2024_BuildingBlockSGSModel,
	author = {Arranz, G. and Ling, Y. and Costa, S. and Goc, K. and Lozano-Dur{\'a}n, A.},
	journal = {Communications Engineering},
	number = {1},
	pages = {127},
	title = {Building-block-flow computational model for large-eddy simulation of external aerodynamic applications},
	volume = {3},
	year = {2024}
}

@article{McKEON_SHARMA_2010, 
    title={A critical-layer framework for turbulent pipe flow}, 
    volume={658}, 
    journal={Journal of Fluid Mechanics}, 
    author={McKeon, B. J. and Sharma, A. S.}, 
    year={2010}, 
    pages={336–382}
}

@article{schmid2002_GlobalStability,
  title={Stability and transition in shear flows. applied mathematical sciences, vol. 142},
  author={Schmid, P. J. and Henningson, D. S. and Jankowski, D. F.},
  journal={Appl. Mech. Rev.},
  volume={55},
  number={3},
  pages={B57--B59},
  year={2002}
}

@article{JLumley_FirstFluidsPODPaper,
  title={The structure of inhomogeneous turbulent flows.},
  author={Lumley, J. L.},
  journal={Atmos. Turbul. Radio Wave Propag},
  pages = "166--178",
  year={1967}
}

@book{Advanced_Mathematical_Methods_for_Scientists_and_Engineers,
    author = {Bender, C. M. and Orszag, S. A.},
    title = {Advanced Mathematical Methods for Scientists and Engineers.},
    year = {1999},
    publisher = {Springer}
}

@article{K41_1941,
	author = {Kolmogorov, A. N.},
	journal = {Dokl. Akad. Nauk SSSR},
	pages = {299-303},
	title = {The local structure of turbulence in incompressible viscous fluid for very large Reynolds numbers},
	year = {1941},
}

@article{Zou_sPCA_2006,
	author = {Zou, H. and Hastie, T. and Tibshirani, R.},
	journal = {Journal of Computational and Graphical Statistics},
	number = {2},
	pages = {265--286},
	title = {Sparse Principal Component Analysis},
	volume = {15},
	year = {2006}
}

@book{Bishop_GMM_2006, 
    title={Pattern recognition and machine learning.}, 
    publisher={Springer New York},
    author={Bishop, C.},
    year={2006}
}

@article{Haller_LCS_2002,
	author = {Haller, G.},
	journal = {Physics of Fluids},
	number = {6},
	pages = {1851-1861},
	title = {Lagrangian coherent structures from approximate velocity data},
	volume = {14},
	year = {2002},
}

@article{tSNE_vandermaaten_JMLR,
  author  = {Maaten, L. van der  and Hinton, G.},
  title   = {Visualizing Data using t-SNE},
  journal = {Journal of Machine Learning Research},
  year    = {2008},
  volume  = {9},
  number  = {86},
  pages   = {2579--2605},
}

@article{UMAP_McInnes2018, 
    year = {2018}, 
    volume = {3}, 
    number = {29}, 
    pages = {861}, 
    author = {McInnes, L. and Healy, J. and Saul, N. and Großberger, L.}, 
    title = {UMAP: Uniform Manifold Approximation and Projection}, 
    journal = {Journal of Open Source Software} 
}

@article{tran2025weakformscientificmachine,
      title={Weak Form Scientific Machine Learning: Test Function Construction for System Identification}, 
      author={Tran, A. and Bortz, D.},
      year={2025},
      journal={arXiv preprint arXiv:2507.03206},
}

@article{Bortz2023_WENDY_BumpFunctions,
	author = {Bortz, D. M. and Messenger, D. A. and Dukic, V.},
	date = {2023/10/05},
	journal = {Bulletin of Mathematical Biology},
	number = {11},
	pages = {110},
	title = {Direct Estimation of Parameters in ODE Models Using WENDy: Weak-Form Estimation of Nonlinear Dynamics},
	volume = {85},
	year = {2023}
}

@article{WANG201944_WSINDy,
	author = {Wang, Z. and Huan, X. and Garikipati, K.},
	journal = {Computer Methods in Applied Mechanics and Engineering},
	pages = {44-74},
	title = {Variational system identification of the partial differential equations governing the physics of pattern-formation: Inference under varying fidelity and noise},
	volume = {356},
	year = {2019},
}

@article{bioinformatics_WSINDy,
	author = {Pantazis, Y. and Tsamardinos, I.},
	journal = {Bioinformatics},
	number = {18},
	pages = {3387-3396},
	title = {A unified approach for sparse dynamical system inference from temporal measurements},
	volume = {35},
	year = {2018},
}

@article{Grigoriev_HisFirstWSINDyPaper,
    author = {Gurevich, D. R. and Reinbold, P. A. K. and Grigoriev, R. O.},
	journal = {Chaos: An Interdisciplinary Journal of Nonlinear Science},
	number = {10},
	pages = {103113},
	title = {Robust and optimal sparse regression for nonlinear PDE models},
	volume = {29},
	year = {2019},
}

@article{WavyWallDataGen_PhysRevFluids.4.034605,
  title = {Streamline segment statistics propagation in inhomogeneous turbulence},
  author = {Rubbert, A. and Albers, M. and Schr\"oder, W.},
  journal = {Phys. Rev. Fluids},
  volume = {4},
  issue = {3},
  pages = {034605},
  numpages = {31},
  year = {2019},
}

@article{Lagemann_WavyWall_OpticalFlow_2024,
	author = {Lagemann, E. and Brunton, S. L. and Lagemann, C.},
	journal = {Proceedings of the Royal Society A: Mathematical, Physical and Engineering Sciences},
	number = {2292},
	pages = {20230798},
	title = {Uncovering wall-shear stress dynamics from neural-network enhanced fluid flow measurements},
	volume = {480},
	year = {2024},
}

@article{Vinuesa2016ConvergenceOfNumSimsOfWallBoundedAndDuctFlows,
	author = {Vinuesa, R. and Prus, C. and Schlatter, P. and Nagib, H. M.},
	journal = {Meccanica},
	number = {12},
	pages = {3025--3042},
	title = {Convergence of numerical simulations of turbulent wall-bounded flows and mean cross-flow structure of rectangular ducts},
	volume = {51},
	year = {2016},
}

@article{Vinuesa_OGDuctARPaper_2014,
	author = {Vinuesa, R. and Noorani, A. and Lozano-Dur{\'a}n, A. and El Khoury, G. K. and Schlatter, P. and Fischer, P. F. and Nagib, H. M.},
	journal = {Journal of Turbulence},
	number = {10},
	pages = {677--706},
	title = {Aspect ratio effects in turbulent duct flows studied through direct numerical simulation},
	volume = {15},
	year = {2014},
}

@article{Kaptanoglu:2020vu,
	author = {Kaptanoglu, A. A. and Morgan, K. D. and Hansen, C. J. and Brunton, S. L.},
	journal = {Physics of Plasmas},
	number = {3},
	pages = {032108},
	title = {Characterizing magnetized plasmas with dynamic mode decomposition},
	volume = {27},
	year = {2020}}

@article{Quarteroni_Manzoni_Vergara_2017, 
    title={The cardiovascular system: Mathematical modelling, numerical algorithms and clinical applications}, 
    volume={26}, 
    journal={Acta Numerica}, 
    author={Quarteroni, A. and Manzoni, A. and Vergara, C.}, 
    year={2017}, 
    pages={365–590}
}

@article{Hudson:1996TurbulenceProduction,
	author = {Hudson, J. D. and Dykhno, L. and Hanratty, T. J.},
	journal = {Experiments in Fluids},
	number = {4},
	pages = {257--265},
	title = {Turbulence production in flow over a wavy wall},
	volume = {20},
	year = {1996}
}

@article{SHAP_Vinuesa,
	author = {Cremades, A. and Hoyas, S. and Deshpande, R. and Quintero, P. and Lellep, M. and Lee, W. J. and Monty, J. P. and Hutchins, N. and Linkmann, M. and Marusic, I. and Vinuesa, R.},
	journal = {Nature Communications},
	number = {1},
	pages = {3864},
	title = {Identifying regions of importance in wall-bounded turbulence through explainable deep learning},
	volume = {15},
	year = {2024}
}

@thesis{Roques2024_DBforRANS,
    author       = {Roques, C.},
    title        = {Aggregation of turbulence models for turbomachinery flows using Bayesian calibration and machine learning},
    school       = {Sorbonne Université},
    year         = {2024},
    language     = {English}
}

@article{Matai2019_ZonalEddyViscosity,
	author = {Matai, R. and Durbin, P. A.},
	journal = {Flow, Turbulence and Combustion},
	number = {1},
	pages = {93--109},
	title = {Zonal Eddy Viscosity Models Based on Machine Learning},
	volume = {103},
	year = {2019}
}

@article{Karman1930_LawOfTheWall,
    author = {Kármán, T. von},
    journal = {Nachrichten von der Gesellschaft der Wissenschaften zu Göttingen, Mathematisch-Physikalische Klasse},
    pages = {58-76},
    title = {Mechanische Aenlichkeit und Turbulenz},
    volume = {1930},
    year = {1930},
}

@article{prandtl1928motion,
  title={über Flüssigkeitsbewegung bei sehr kleiner Reibung},
  journal = {Verhandl III, Intern. Math. Kongr.},
  author={Prandtl, L.},
  year={1904},
  volume = {2},
  pages = {484-491}
}

@article{BuckinghamPi,
	author = {Buckingham, E.},
	journal = {Phys. Rev.},
	pages = {345 - 376},
	title = {On physically similar systems; Illustrations of the use of dimensional equations},
	volume = {4},
	year = {1914},
}

@inbook{Sterrett_PhysicallySimilarSystems,
	author = {Sterrett, S. G.},
	booktitle = {Springer Handbook of Model-Based Science},
	pages = {377--411},
	publisher = {Springer International Publishing},
	title = {Physically Similar Systems - A History of the Concept},
	year = {2017},
}

@article{Brunton2016pnas,
	author = {Brunton, S. L. and Proctor, J. L. and Kutz, J. N.},
	journal = {Proceedings of the National Academy of Sciences},
	number = {15},
	pages = {3932--3937},
	title = {Discovering governing equations from data by sparse identification of nonlinear dynamical systems},
	volume = {113},
	year = {2016}
}

@article{Rudy2017sciadv,
	author = {Rudy, S. H. and Brunton, S. L. and Proctor, J. L. and Kutz, J. N.},
	journal = {Science Advances},
	number = {e1602614},
	title = {Data-driven discovery of partial differential equations},
	volume = {3},
	year = {2017}
}

@article{WeakSINDY_GalerkinBasedODERegression,
    author = {Messenger, D. A. and Bortz, D. M.},
    title = {Weak SINDy: Galerkin-Based Data-Driven Model Selection},
    journal = {Multiscale Modeling \& Simulation},
    volume = {19},
    number = {3},
    pages = {1474-1497},
    year = {2021},
}

@article{IntegralSINDy,
  title = {Sparse model selection via integral terms},
  author = {Schaeffer, H. and McCalla, S. G.},
  journal = {Phys. Rev. E},
  volume = {96},
  issue = {2},
  pages = {023302},
  numpages = {7},
  year = {2017},
}

@article{WSINDyForPDEs,
	author = {Messenger, D. A.  and Bortz, D. M. },
	journal = {Journal of Computational Physics},
	pages = {110525},
	title = {Weak SINDy for partial differential equations},
	volume = {443},
	year = {2021}
}

@article{EnsembleSINDy,
    author = {Fasel, U. and Kutz, J. N. and Brunton, B. W. and Brunton, S. L. },
    title = {Ensemble-SINDy: Robust sparse model discovery in the low-data, high-noise limit, with active learning and control},
    journal = {Proceedings of the Royal Society A: Mathematical, Physical and Engineering Sciences},
    volume = {478},
    number = {2260},
    pages = {20210904},
    year = {2022},
}

@article{CallahamDB,
	author = {Callaham, J. L. and Koch, J. V. and Brunton, B. W. and Kutz, J. N. and Brunton, S. L.},
	date = {2021/02/15},
	journal = {Nature Communications},
	number = {1},
	pages = {1016},
	title = {Learning dominant physical processes with data-driven balance models},
	volume = {12},
	year = {2021},
}

@article{ROWLEY_DMD, 
    title={Spectral analysis of nonlinear flows}, 
    volume={641}, 
    journal={Journal of Fluid Mechanics}, 
    author={Rowley, C. W. and Mezi\'c, I. and Bagheri, S. and Schlatter, P. and Henningson, D.S.}, 
    year={2009}, 
    pages={115–127}
}

@article{lagemann2024impact,
  title={Impact of Reynolds number on the drag reduction mechanism of spanwise travelling surface waves},
  author={Lagemann, E. and Albers, M. and Lagemann, C. and Schr{\"o}der, W.},
  journal={Flow, Turbulence and Combustion},
  volume={113},
  number={1},
  pages={27--40},
  year={2024},
  publisher={Springer}
}

@article{Lozano_Duran_Bae_2023, title={Machine learning building-block-flow wall model for large-eddy simulation}, volume={963}, DOI={10.1017/jfm.2023.331}, journal={Journal of Fluid Mechanics}, author={Lozano-Durán, A. and Bae, H. J.}, year={2023}, pages={A35}}

@article{JHLTBD_2008,
  title={A public turbulence database cluster and applications to study Lagrangian evolution of velocity increments in turbulence},
  author={Li, Y. and Perlman, E. and Wan, M. and Yang, Y. and Meneveau, C. and Burns, R. and Chen, S. and Szalay, A. and Eyink, G.},
  journal={Journal of Turbulence},
  volume={9},
  number={31},
  year={2008},
  publisher={Nature Publishing Group UK London}
}

@article{StreaksSpotsTransistion_Zaki2013,
	author = {Zaki, T. A. },
	journal = {Flow, Turbulence and Combustion},
	number = {3},
	pages = {451--473},
	title = {From Streaks to Spots and on to Turbulence: Exploring the Dynamics of Boundary Layer Transition},
	volume = {91},
	year = {2013},
}

@inproceedings{lagemann2025hydrogym_a,
  title={HydroGym: A Reinforcement Learning Platform for Fluid Dynamics},
  author={Lagemann, C. and Paehler, L. and Callaham, J. and Mokbel, S. and Ahnert, S. and Lagemann, K. and Lagemann, E. and Adams, N. and Brunton, S. L.},
  booktitle={7th Annual Learning for Dynamics \& Control Conference},
  pages={497--512},
  year={2025},
  organization={PMLR}
}

@article{lagemann2025hydrogym_b,
  title={The HydroGym Reinforcement Learning Platform for Fluid Dynamics},
  author={Lagemann, C. and Mokbel, S. and Gondrum, M. and R{\"u}ttgers, M. and Wang, Y. and Suárez, P. and Paehler, L. and Bezgin, D. A. and Buhendwa, A. B. and Callaham, J. and Ahnert, S. and Zolman, N. and Shao, X. and Loiseau, J. C. and Adams, N. and Meinke, M. and Schr{\"o}der, W. and Lagemann, K. and Lagemann, E. and Vinuesa, R. and Brunton, S. L.},
  journal={arXiv preprint arXiv:2512.17534},
  year={2025}
}

@article{quarteroni2017integrated,
  title={Integrated heart—coupling multiscale and multiphysics models for the simulation of the cardiac function},
  author={Quarteroni, A. and Lassila, T. and Rossi, S. and Ruiz-Baier, R.},
  journal={Computer Methods in Applied Mechanics and Engineering},
  volume={314},
  pages={345--407},
  year={2017},
  publisher={Elsevier}
}

@article{park2025kinetic,
  title={Kinetic turbulence drives MHD equilibrium change via 3D reconnection},
  author={Park, J. Y. and Yoon, Y. D. and Hwang, Y. S.},
  journal={Nature},
  volume={644},
  number={8075},
  pages={59--63},
  year={2025},
  publisher={Nature Publishing Group UK London}
}

@article{bodnar2025foundation,
  title={A foundation model for the Earth system},
  author={Bodnar, C. and Bruinsma, W. P. and Lucic, A. and Stanley, M. and Allen, A. and Brandstetter, J. and Garvan, P. and Riechert, M. and Weyn, J. A. and Dong, H. and others},
  journal={Nature},
  volume={641},
  number={8065},
  pages={1180--1187},
  year={2025},
  publisher={Nature Publishing Group UK London}
}

@article{zeng2020complex,
  title={Complex reaction processes in combustion unraveled by neural network-based molecular dynamics simulation},
  author={Zeng, J. and Cao, L. and Xu, M. and Zhu, T. and Zhang, J. Z. H.},
  journal={Nature communications},
  volume={11},
  number={1},
  pages={5713},
  year={2020},
  publisher={Nature Publishing Group UK London}
}

@inproceedings{JHLTBD_2007, 
    author={Perlman, E. and Burns, R. and Li, Y. and Meneveau, C.},
    title = {Data exploration of turbulence simulations using a database cluster}, 
    year = {2007}, 
    publisher = {Association for Computing Machinery}, 
    booktitle = {Proceedings of the 2007 ACM/IEEE Conference on Supercomputing}, 
    articleno = {23}, 
    series = {SC '07} 
}
 }
 \end{spacing}

\newpage

\appendix

\section*{Supplementary Information}

\SIsection{Turbulent Boundary Layer Additional Error Metrics}
\label{sec: Appendix_ErrorMetric}
We include these additional calculations of the adjusted Rand index~\cite{hubert1985comparing} (ARI) and normalized mutual information~\cite{vinh2010nmi} (NMI) to complement our evaluation of the noise robustness of both weak and finite difference based methods. These metrics both specifically evaluate the accuracy of the regions that we cluster together, matching GMM labels from a noisy trial against those from a no-noise baseline trial at each point in the domain to measure their agreement. While these metrics only evaluate one part of the full dominant balance algorithm (the clustering step), we find this useful to illustrate the rapid qualitative breakdown of the partitions identified by the pointwise method, whereas the median ARI and NMI for the weak method remains stable through $100\%$ noise. 
\begin{figure}[h!]
    \centering
    
    \includegraphics[width=\textwidth]{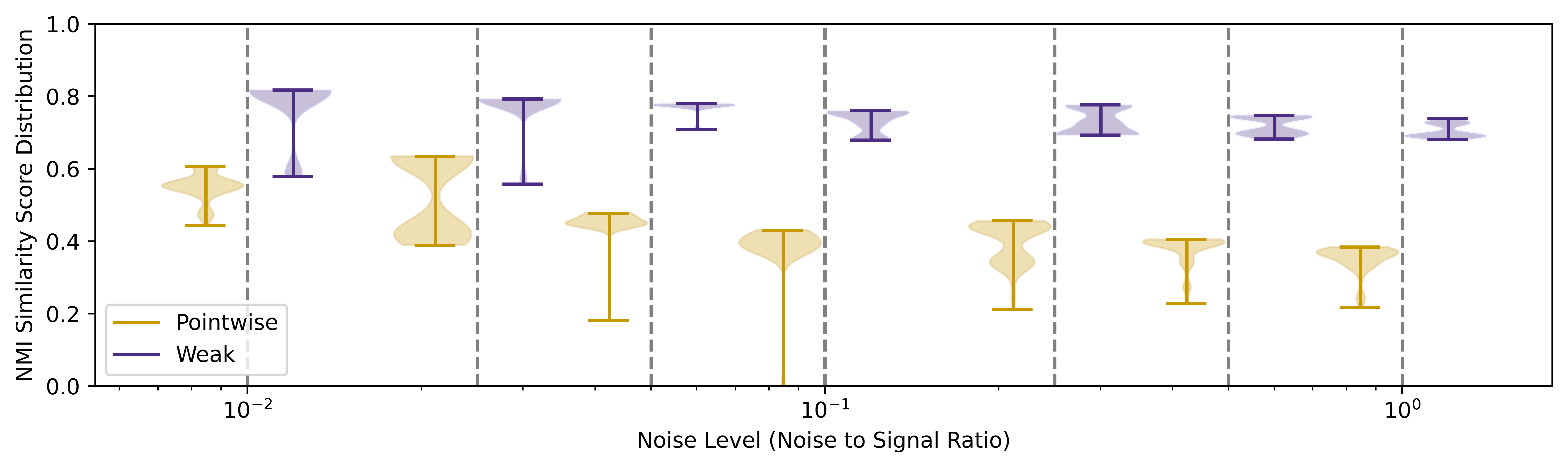}
    
    \vspace{0.5cm} 
    
    \includegraphics[width=\textwidth]{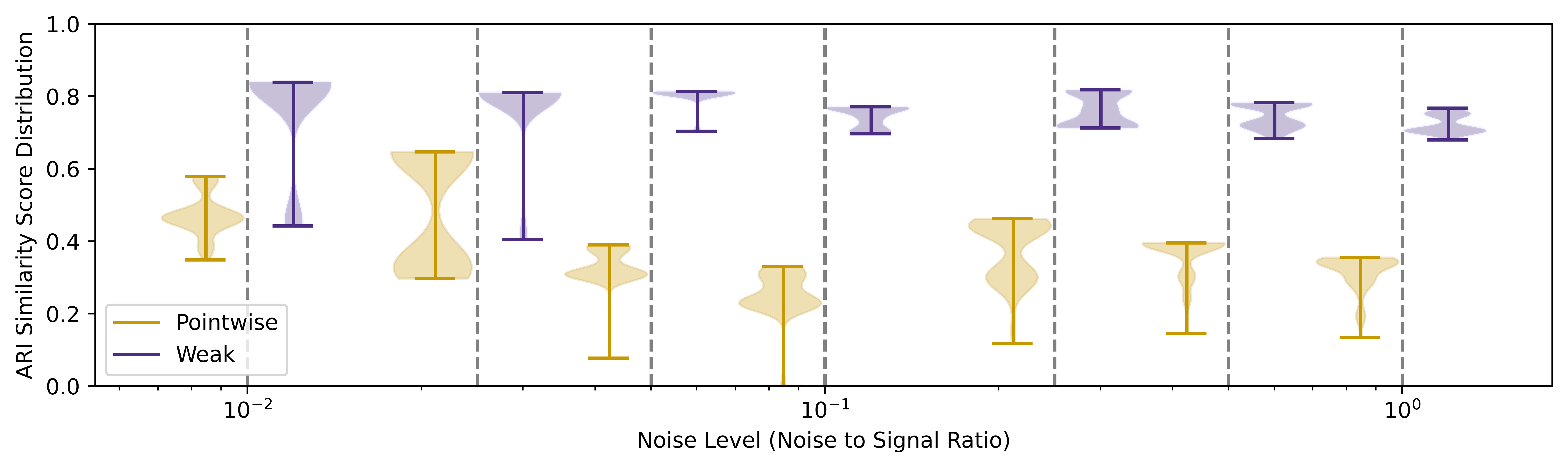}
    
    \caption{\textbf{Two label-permutation-invariant clustering metrics confirm the noise robustness of the weak formulation across the full range from 1\% to 100\% noise.} Distributions of two label-permutation-invariant clustering similarity scores---normalised mutual information (NMI,top) and adjusted Rand index (ARI, bottom)---computed between the noise-free baseline decomposition and each noisy trial of the flat-plate boundary layer benchmark (Fig.~\ref{Fig:2QuantComparison}). Higher values indicate better agreement; both metrics range from 0 (random labelling) to 1 (perfect agreement). Violin shapes show the distribution across 100 independent trials at each noise level, with whiskers marking the minimum, median, and maximum. The weak formulation (purple) maintains median NMI $\approx 0.7$ and median ARI $\approx 0.75$ across the entire noise range from 1\% to 100\%, whereas the pointwise baseline (yellow) degrades to median NMI $\approx 0.35$ and median ARI $\approx 0.30$ by 5\% noise. Unlike the $\mathcal{L}_0$ metric used in the main text, NMI and ARI are invariant to arbitrary permutation of cluster labels by construction and weight cluster-boundary disagreements differently. The consistent ordering of methods across all three metrics confirms that the reported robustness gap reflects a real difference in classification fidelity rather than a metric- specific artifact.}
    \label{Fig:NMI_ARI_Result}
\end{figure}

\SIsection{Data-Driven Dominant Balance}
\label{SI:DominantBalance}
Governing equations derived from first principles have proven to be a central tool in accurately predicting the complex behaviors of physical systems in science and engineering. 
Although many systems often exhibit an immense diversity of behaviors driven by either chaotic or multi-scale physics, the majority of their evolution is frequently determined by only a few important dominant processes.  
Under this assumption, only a subset of leading-order terms in the governing equation may be used to approximate the local dynamics in a self-similar region of the domain producing ``balance models''. 
The data-driven dominant balance (DD-DB) method uses unsupervised learning to identify these balance regions within a particular dynamical system, isolating the critical mechanisms that drive behaviors and form structures across operating regimes ~\cite{CallahamDB}.

The problem set up begins with a governing partial differential equation in a field variable $u$, where the dynamics may be linearly separated into a sum of individual terms:
\begin{equation}
\label{eqn: WeakDBGovEqn}
\mathcal{N}(u) = \sum_{i=1}^{k} f_1(u) + f_2(u) + \dots + f_k(u) = 0.
\end{equation}
For example, in the paper by \emph{Callaham et al.}~\cite{CallahamDB}, the streamwise component of the two-dimensional Reynolds averaged Navier-Stokes (RANS) equations take the form:
\begin{equation}
\overline{U} \ \overline{U}_x + \overline{V} \ \overline{U}_y = -\frac{\overline{p}_x}{\rho} + \nu(\overline{U}_{xx} + \overline{U}_{yy}) - (\overline{u'^2}_x + \overline{u'v'}_y).
\label{eqn: RANS_1} 
\end{equation}
Classical boundary layer theory indicates that the physical space is segmented into regions where subsets of these terms are active and in balance, while others are approximately zero.  For example, in the far field only a balance of the streamwise advection $\bar{U}\bar{U}_x$ and pressure gradient $-\bar{p}_x/\rho$ terms are relevant, whereas the near-wall region involves a balance between the viscous $\nu(\overline{U}_{xx} + \overline{U}_{yy})$ and Reynolds stress term $\overline{u'v'}_y$.

Instead of the traditional scaling or Buckingham Pi approaches, the DD-DB class of methods use \emph{unsupervised learning}  to apply the dominant balance heuristic and identify these regions of the domain. First, the method reconstructs each term in the governing Eq. \ref{eqn: WeakDBGovEqn}, and defines a coordinate system where each term defines a dimension within a new ``equation space'' frame of reference, resulting in the following vector:
\begin{equation}
    \label{eqn: eqnSpaceVector}
    \mathcal{V} = [f_1(u), f_2(u), \dots, f_k(u)]^T.
\end{equation}
Each point in the spatio-temporal domain is now directly mapped into the higher-order equation space domain where the data has been recast to be representative of the varying scales of dynamics that drive the system. 
Points in equation space produce manifolds as the varying governing physical mechanisms transition and separate between distinct regimes in the system which allows for 1.) applying unsupervised learning to cluster regions of similar dynamics and 2.) subspace identification to determine the directions of highest variance in equation space. 
In particular, ~\cite{CallahamDB} uses Gaussian Mixture Modelling (GMM)~\cite{Bishop_GMM_2006} and sparse principal component analysis (sPCA)~\cite{Zou_sPCA_2006} to extract a spatial decomposition and identify active terms in each region. However, other approaches to eliminate the presupposed Gaussian form of the data by GMM or to avoid the use of a global sparsity threshold as in sPCA may be useful depending on the unique structure within a given equation space.

The final balance models delivered by DD-DB serve as a reduced representation of the local dynamics within a region of interest. These are primarily viewed by the authors of~\cite{CallahamDB} as active subspaces in equation space, capturing the axes of highest variance to isolate only the relevant dynamics for state propagation. However, dominant balance may also offer insight into the multi-regime nature of a physical system and the distinct dynamics that give rise to often rich and intricate behaviors. 
Particularly in fluid dynamics, these balance models may provide a condensed summary of the mechanisms actively driving flow behavior and the structures that result from these mechanisms, as we explore in our paper above. 

\end{document}